\documentclass[aps,pra,epsfig,twocolumn,showpacs]{revtex4}
\usepackage{graphicx}
\usepackage{dcolumn}
\usepackage{amsthm}

\usepackage{mathrsfs}
\usepackage{bm}
\usepackage{psfig}
\usepackage{amsmath}
\usepackage{amssymb}
\input{Qcircuit}

\newcommand{\iden}{1 \hspace{-1.0mm}  {\bf l}}

\newcommand{\ncd}{\newcommand}
\ncd{\QC}{$\mbox{QC}_{\cal{C}}$}
\ncd{\QCpr}{${\mbox{QC}_{\cal{C}}}^\prime\;$}
\ncd{\QCns}{$\mbox{QC}_{\cal{C}}$}
\ncd{\QCprns}{${\mbox{QC}_{\cal{C}}}^\prime$}
\ncd{\cskN}{{|\phi_{\{\kappa\} } \rangle}_{{\cal{C}}_N}}
\ncd{\cskNpr}{{|\phi_{\{\kappa^\prime\} } \rangle}_{{\cal{C}}_N}}
\ncd{\cskNtil}{{|\phi_{\{\tilde{\kappa} \} }
\rangle}_{{\cal{C}}_N}}
\ncd{\csk}{{|\phi_{\{\kappa\} }
\rangle}_{\cal{C}}}
\ncd{\csktil}{{|\phi_{\{\tilde{\kappa} \} }
\rangle}_{\cal{C}}}
\ncd{\cskf}{|\phi_{\{\kappa\} }
\rangle_{\cal{C}}}
\ncd{\csktilf}{|\phi_{\{\tilde{\kappa} \} }
\rangle_{\cal{C}}}
\ncd{\bracsk}{\mbox{}_{\cal{C}}\langle\phi_{\{\kappa\} }|}
\ncd{\bracsktil}{\mbox{}_{\cal{C}}\langle\phi_{\{\tilde{\kappa} \}
}|} \ncd{\nbracsk}{\mbox{}_{\cal{C}}\langle\phi_{\{\kappa\} }}
\ncd{\nbracsktil}{\mbox{}_{\cal{C}}\langle\phi_{\{\tilde{\kappa}
\} }} \ncd{\cs}{|\phi \rangle_{\cal{C}}\;} \ncd{\csns}{|\phi
\rangle_{\cal{C}}}
\ncd{\nbgh}{\text{nghb}} \ncd{\Sab}{S^{ab}}
\ncd{\Sba}{S^{ba}}
\ncd{\ds}{\displaystyle} \ncd{\ovl}{\overline}

\newtheorem{Theorem}{Observation}

\begin{document}
\setlength{\hoffset}{-0.2 cm}


\title{Quantum information processing with noisy cluster states}
\author{M. S. Tame$^1$, M. Paternostro$^1$, M. S. Kim$^1$, V. Vedral$^{2,3}$}
\affiliation{$^1$School of Mathematics and Physics, The Queen's University, Belfast, BT7 1NN, UK}
\affiliation{$^2$Institute of Experimental Physics, University of Vienna, Boltzmanngasse 5, 1090 Vienna, Austria\\
$^3$The School of Physics and Astronomy, University of Leeds, Leeds, LS2 9JT, UK}

\date{\today}
\begin{abstract}
We provide an analysis of basic quantum information processing protocols under the effect of intrinsic non-idealities in cluster states. 
These non-idealities are based on the introduction of randomness in the entangling steps that create the cluster state and are motivated by the unavoidable imperfections faced in creating entanglement using condensed-matter systems. 
Aided by the use of an alternative and very efficient method to construct cluster state configurations, which relies on the concatenation of fundamental cluster structures, we address quantum state transfer and various fundamental gate simulations through noisy cluster states. We find that a winning strategy to limit the effects of noise, is the management of small clusters processed via just a few measurements. Our study also reinforces recent ideas related to the optical implementation of a one-way quantum computer. 
\end{abstract}

\pacs{03.67.Lx,03.67.Mn,42.50.Dv}

\maketitle

\section{Introduction}

The one-way quantum computer (\QC)~\cite{RBH,rb1} is an intriguing model for quantum computation (QC) which represents an alternative viewpoint from the canonical quantum circuit model~\cite{nielsenchuang}. The idea behind \QC~schemes is that single-qubit rotations and two-qubit entangling operations can be effectively simulated by the use of an entangled resource, the so-called {\it cluster state}~\cite{rb2} shared by a group of qubits. Quantum Information Processing (QIP) is carried out by an adaptive set of single-qubit measurements, specifically designed for the particular simulation desired. Until present, several studies have been undertaken in relation to the generation, characterization and detection of cluster states, as well as more general graph-states~\cite{bloch,rb2,mouraalves,toth}. In addition, it has been recognized by Nielsen~\cite{nielsenprl} that cluster states could be used in order to improve the efficiency of the scheme for linear optics QC suggested by Knill, Laflamme and Milburn~\cite{KLM}. Very recently, we have seen the first experimental evidence of one-way computation, with the realization of a quantum search algorithm achieved through an all-optical setup~\cite{vlatko}. The class of cluster states and their related protocols for QC most certainly deserve a deeper analysis and understanding. For example, it is still not clear if the one-way model is more robust against the effects of external sources of noise and imperfection than the quantum circuit model. Several attempts with regards to this have shown that any Markovian noise in the one-way model can be mapped onto non-Markovian imperfections affecting the corresponding quantum circuit being simulated~\cite{nielsendawson}. Thus, one can take advantage of already existing theorems, valid for the quantum circuit model~\cite{terhal}, to find proper thresholds for fault-tolerant cluster state computation~\cite{nielsendawson}. However we have recently introduced an experimentally motivated model for imperfect cluster state generation, which paves the way toward the study of the performances of one-way computation exposed to realistic sources of non-ideality~\cite{noiarchivio}. Several important results were presented, which showed how the effectiveness of simulations are affected by randomness introduced into the structure of the cluster state being used. This led us to search for schemes of cluster state QC based on the use of smaller cluster configurations than originally devised~\cite{RBH}. 
In this paper we analyze the effects of the model for noise on various key QIP protocols, while at the same time using a {\it concatenation} technique for the construction of compact configurations for cluster state QC. Elementary cluster basic building blocks ($BBB$'s) will be introduced, such that they can be concatenated to form larger structures for specific circuit simulations. With this alternative technique, one may bypass the cumbersome stabilizer formalism approach~\cite{RBH} and quickly construct more economical circuit simulations using cluster states.

The paper is structured as follows: In Section~\ref{Intro}, we review the basic properties of cluster states and introduce the notation which will be used throughout the rest of the paper. The concatenation approach will be discussed in Section~\ref{tessellation} and compared to the standard approach of constructing cluster state configurations. This alternative viewpoint will then be adopted throughout the rest of our analysis. Section~\ref{modellonostro} introduces our model for imperfect cluster state generation, where we review the fundamental properties of {\it noisy cluster states} found in~\cite{noiarchivio}. The implications of the profound differences between the underlying ideal and non-ideal cluster state structures, supplies us with strong motivations to seek a deeper analysis of QIP protocols using noisy cluster states. 
Section~\ref{informationflow} is devoted to the transfer of quantum information through a linear cluster of arbitrary length and we see how the average transfer fidelity depends on the dimension of the cluster, with this rapidly becoming smaller than the best classical fidelity. Such a result is seen as an effect of spreading pre-existing noise by the increasing number of single-qubit measurements required in order to process quantum information encoded in a cluster state. We call this phenomenon {\it inheritance of noise} by the surviving qubits in a cluster after a given set of measurements. This effect is further investigated in Section~\ref{computation}, where we address the simulation of single-qubit and two-qubit entangling quantum gates in noisy cluster states. Our analysis includes the basic gates, such as the controlled-{\sf NOT} ({\sf CNOT})~\cite{nielsenchuang} and general single-qubit rotations. The results of our investigation lead to the conclusion that QIP in the presence of noise is viable only for registers of just a few qubits. This pushes us toward research of more economical strategies for gate simulation~\cite{noiarchivio} than those suggested in the original one-way model~\cite{RBH}. Finally, Section~\ref{remarks} summarizes our results.

\section{Background for cluster states}
\label{Intro}
We begin our investigation by reviewing some of the elements of the \QC~model. The stabilizer formalism will be used extensively in this Section. This is a necessity, as we will want to compare it with the more advantageous cluster state concatenation approach.

A cluster state $\csk$ is a pure multipartite entangled state of qubits
positioned at specific sites of a lattice structure known as the
cluster ${\cal{C}}$. The complete characterization of cluster states is the starting point 
in understanding of how the \QC~model works. A cluster state $\csk$ is formally defined as the eigenstate of the set of Hermitian operators 
\begin{equation}
    \label{BasCorr}
    K^{(a)} =  \sigma_x^{(a)}
    \bigotimes\limits_{b \in \nbgh(a) \cap {\cal{C}}}
    \sigma_z^{(b)}
\end{equation}
such that
 $K^{(a)}\csk  = (-1)^{\kappa_a}\csk$~\cite{RBH,rb1}.
Here, $\sigma_{x}$ ($\sigma_{z}$) is the $x$ ($z$) Pauli matrix, while each 
$K^{(a)}$ acts on the qubit occupying site $a\in{\cal{C}}$ and 
on any other qubit occupying a neighboring lattice site $b\in\nbgh(a)\cap{\cal{C}}$. They are {\it correlation operators} which form a complete set of $|{\cal{C}}|$ independent and
commuting observables for the qubits within ${\cal{C}}$. Thus, a cluster state $\csk$ is
completely specified by the set of numbers $\{\kappa\}:=\{\kappa_a\in\{0,1\}|\,a\in{\cal{C}}\}$.
 For example, a different cluster
state ${|\phi_{\{\kappa^{\prime}\} } \rangle}_{\cal{C}}$ belonging
to the same physical cluster ${\cal{C}}$ ({\it i.e.} the same lattice
configuration), is completely described by the elements in the set 
$\{ \kappa^\prime \} := \{\kappa_a^\prime
\in \{0,1\}|\, a \in {\cal{C}}\}$. All the cluster states of an $N$ qubit cluster ${\cal C}_N$ correspond to the $2^{| \cal{C} |}$ possible combinations of
the elements in $\{ \kappa \}$. They are mutually
orthogonal and form a basis in the $2^N$-dimensional Hilbert space of the cluster. These states are also all equivalent under the application of $\sigma_z^{(a)}$ on individual qubits. In this paper, we will often use the simplified notation $|\phi \rangle_{{\cal{C}}}:=|\phi_{ \{ \kappa \} }\rangle_{{\cal{C}}}$, with $\{\kappa \}=\{0,\,\forall a\in{\cal{C}}\}$.
This particular cluster state is generated by first preparing 
the product state $|+\rangle_{\cal{C}}= \bigotimes_{a
\in {\cal{C}}} |+\rangle_{a}$ of the qubits at all the sites $a$ of ${\cal{C}}$. In our notation, $|\pm\rangle_a=(1/{\sqrt 2})(|0\rangle\pm |1\rangle)_a$ are the eigenstates of $\sigma^{(a)}_x$ and $\ket{0}_{a}$ ($\ket{1}_{a}$) is the eigenstate of $\sigma^{(a)}_{z}$ corresponding to the $+1$ ($-1$) eigenvalue. We then apply the unitary transformation $S^{({\cal{C}})}=\prod_{\langle a,b\rangle}\Sab$ to this initial state, where $\langle a,b\rangle:=\{a,b\in{\cal{C}}|b-a\in\gamma_d\}$ and $\gamma_1=\{1\}$, $\gamma_2=\{(1,0)^T,(0,1)^T\}$, $\gamma_3=\{(1,0,0)^T,(0,1,0)^T, (0,0,1)^T \}$ for the respective dimension
$d$ of the cluster being used. Each operator $\Sab$ can be described by the controlled phase gate
\begin{equation}
    \label{Sabdef}
    \Sab= |0 \rangle_a \langle 0| \otimes {\iden}^{(b)} + |1 \rangle_a \langle 1| \otimes \sigma_z^{(b)}
\end{equation}
and is an entangling operation between the qubits at
sites $a$ and $b$ of the cluster ${\cal{C}}$. It is important to
note that all $\Sab$'s mutually commute and therefore the time required for the generation of a $N$-qubit cluster state is independent from $N$~\cite{RBH}. 
The state generated by the action of $S^{({\cal{C}})}$
on $|+\rangle_{\cal{C}}$ is found to be~\cite{RBH}
\begin{equation}\label{entgl}
S^{({\cal{C}})} |+\rangle_{\cal{C}}\equiv\prod\limits_{<a,b>}
\Sab \bigotimes\limits_{a \in {\cal{C}}}
    |+\rangle_{a}=|\phi \rangle_{{\cal{C}}}.
\end{equation}
An Ising type interaction between the qubits in a square lattice produces a time-evolution operator having the form of $S^{({\cal{C}})}$. 


\subsection{$QC_{\cal C}$ Operation}
\label{operation}

Let us assume we want to carry out a unitary operation $U_g$ associated with
a particular quantum gate $g$ which acts on an unknown input
state $|\psi_{\rm{in}}\rangle$ of $n$ logical qubits. The idea behind the \QC~model is to use a cluster of $N$ qubits in a particular physical configuration,
${\cal{C}}(g)$. In order to better understand how the 
$U_g$ operation on ${\cal{C}}(g)$ takes place, it is convenient to consider 
the cluster as partitioned into three sections. First, we consider an $n$-qubit input section 
${\cal{C}}_I(g)$ which encodes the input state $|\psi_{\rm{in}}
\rangle$ of the register. We then have a body section ${\cal{C}}_M(g)$ and an $n$-qubit output section ${\cal{C}}_O(g)$ for the
read-out of $U_g$. The three sections have no mutual overlap and taken altogether, they reconstruct the structure of the cluster ({\it i.e.}
${\cal{C}}_\alpha(g)\cap{\cal{C}}_\beta(g)=\emptyset,\,\cup_{\alpha}{\cal{C}}_\alpha(g)={\cal{C}}(g)$,
where $\alpha,\beta\in \{I,M,O \}$ and $\alpha \neq \beta$). We begin by preparing $|\psi_{\rm{in}}\rangle$ on the
cluster qubits in ${\cal{C}}_I(g)$ and denote this state as $|\psi_{\rm{in}}
\rangle_{{\cal{C}}_I(g)}$. It satisfies
\begin{equation}
|\psi_{\rm{in}} \rangle_{{\cal{C}}_I(g)}=
P^{{\cal{C}}_I(g)}_{Z_{,\bf{\{Z
\}}}}[ \alpha_i ] \bigotimes\limits_{j =1}^n |+\rangle_{j}
\end{equation}
with $\{ {\bf Z
} \}:= \{ {\bf z}_i|i \in 2^n \}$ and the multi-qubit projector
$P^{{\cal{C}}_I(g)}_{Z_{,{\bf{ \{ Z \}
}}}} [ \alpha_i ]=\sum_{i=1}^{2^n} \alpha_i
P^{{\cal{C}}_I(g)}_{Z_{,{\bf{z}}_i}}$.
Here, we have introduced the single ${\bf{z}}_i$ state projector
\begin{equation}
P^{{\cal{C}}_I(g)}_{Z_{,{\bf{z}}_i}}=\bigotimes\limits_{j=1}^n
\frac{1+(-1)^{s^{{\bf z}_i}_j} \sigma_z^{[j]}}{2}
\end{equation}
which projects each qubit in ${\cal{C}}_I(g)$ into the state
$|s_j^{{\bf z}_i}\rangle_{j}$,  with $s_j^{{\bf z}_i}$ as the value of the j-th binary 
digit of the integer ${\bf z_i}$ and $[ j ]$ corresponding to an operation
on the logical qubits of the cluster. It
should be made clear that the multi-qubit projector
$P^{{\cal{C}}_I(g)}_{Z_{,{\bf{ \{ Z \}
}}}} [ \alpha_i ]$ does not correspond to measurements which are performed in reality, but is
fictitiously introduced in order to relate the logical input state
$|\psi_{\rm{in}} \rangle$ to any possible state present on the
input qubits of the cluster~\cite{RBH}. In other words, ${\cal{C}}_I(g)$ is
analogous to the register in the quantum circuit model and $|\psi_{\rm{in}}
\rangle_{{\cal{C}}_I(g)}$ is an arbitrary superposition of computational states
$|\psi_{\rm{in}} \rangle_{{\cal{C}}_I(g)}= \sum^{2^n}_{i=1}\alpha_i
|{\bf z}_i \rangle,\,(\sum^{2^n}_{i=1}|\alpha_i|^2=1)$.
After this preparation of $|\psi_{\rm{in}}
\rangle_{{\cal{C}}_I(g)}$, the cluster is entangled by the operation $S^{{\cal{C}}(g)}$ given in eq. (\ref{entgl}). A measurement pattern ${\cal{M}}^{{\cal{C}}_M(g)}$ is then applied to the {\it body section}
of the cluster ${\cal{C}}_M(g)$. The 
pattern is specified by a set of vectors $\vec{r}_a$ ($a\in {\cal{C}}_M(g)$) defining the bases in the Bloch sphere of a set of single-qubit measurements to be performed in ${\cal{C}}_M(g)$. We denote 
    $\{{s}\}=\left\{{s}_a \in \{0,1\} \,|\,\, a \in
        {{\cal{C}}_M(g)} \right\}$ as the set of measurement outcomes
obtained after ${\cal{M}}^{{\cal{C}}_M(g)}$ is applied. Performing a measurement pattern on the body section of a cluster is 
formally equivalent to applying the projector
\begin{equation}
\label{proiettore}
    P^{{\cal{C}}_M(g)}_{{\{{s}\}}}({\cal{M}})=
    \bigotimes_{k \in {{\cal{C}}_M(g)}}
    \frac{1+(-1)^{{s}_k} \vec{r}_k\cdot
      \vec{\sigma}^{(k)} }{2}
\end{equation}
to the initial state. It is easy to see that $S^{{\cal{C}}(g)}$ and 
$P^{{\cal{C}}_I(g)}_{Z_{,{\bf{ \{ Z \}
}}}} [ \alpha_i ]$ commute. With $P^{{\cal{C}}_M(g)}_{{\{{s}\}}}({\cal M})$ acting only on the body section ${\cal{C}}_M(g)$ of the cluster, the state of
the entire cluster ${\cal{C}}(g)$ after this procedure can be written as
\begin{equation}
\label{ancorano}
P^{{\cal{C}}_I(g)}_{Z_{,{\bf{ \{ Z \}
}}}} [ \alpha_i ]\underbrace{P^{{\cal{C}}_M(g)}_{{\{{s}\}}}({\cal{M}})
\underbrace{S^{({\cal{C}}(g))}|+\rangle_{{\cal{C}}(g)}}_{| \phi
\rangle_{{\cal{C}}(g)}}}_{|\psi \rangle_{{\cal{C}}(g)}}.
\end{equation}
While the body qubits have been removed by the set of measurements in 
${\cal{M}}^{{\cal{C}}_M(g)}$, Eq.~(\ref{ancorano}) still contains degrees of freedom in the input section. To eliminate them and obtain just the cluster qubits of ${\cal C}_{O}(g)$, we perform 
measurements in the $\sigma_x$-eigenbasis on each qubit in ${\cal C}_{I}(g)$. That is 
\begin{equation}
\label{orasi}
\ket{\psi_{\rm out}}_{{\cal C}_{O}(g)}=P^{{\cal{C}}_I(g)}_{{\{{s}\}}}(X)
P^{{\cal{C}}_I(g)}_{Z_{,{\bf{ \{ Z \}
}}}} [ \alpha_i ]{| \psi \rangle_{{\cal{C}}(g)}}.
\end{equation}
\newline
A theorem central to the cluster state model~\cite{RBH} states that if $|\psi\rangle_{{\cal{C}}(g)}$ satisfies the set of eigenvalues equations  
        \begin{eqnarray}
        \label{gatecheck}
            \sigma_x^{{\cal{C}}_I(g),j} \left( U_g \sigma_x^{(j)}
            U_g^\dagger \right)^{{\cal{C}}_O(g)}
            |\psi\rangle_{{\cal{C}}(g)} &=& {(-1)}^{\lambda_{x,j}}
            |\psi\rangle_{{\cal{C}}(g)},\nonumber\\
            \sigma_z^{{\cal{C}}_I(g),j} \left( U_g \sigma_z^{(j)}
            U_g^\dagger \right)^{{\cal{C}}_O(g)}
            |\psi\rangle_{{\cal{C}}(g)} &=& {(-1)}^{\lambda_{z,j}}
            |\psi\rangle_{{\cal{C}}(g)},\nonumber \\
        \end{eqnarray}
the input state $\ket{\psi_{\rm in}}$ and the output $\ket{\psi_{\rm out}}$ are related via
\begin{equation}
    \label{acts}
    |\psi_{\text{\rm{out}}}\rangle = U_g U_\Sigma
     \,|\psi_{\text{\rm{in}}} \rangle,
\end{equation}
where $U_\Sigma$ is a byproduct operator, acting locally on the logical qubits, given by
\begin{equation}
    \label{ByProp}
    U_\Sigma=
    \bigotimes_{j=1}^{n}(\sigma_z^{[j]})^{{s}_j^x+\lambda_{x,j}}
    (\sigma_x^{[j]}) ^{\lambda_{z,j}}.
\end{equation}
Thus, the realization of the appropriate measurement pattern effectively simulates the action of the gate $g$. However, we must counteract the unwanted effect of the byproduct operator which appears in the simulation. If $g$ belongs to the Clifford group~\cite{clifford}, the propagation of $U_\Sigma$ through $U_g$ in Eq.~(\ref{acts}) leaves $U_g$ unaffected~\cite{RBH}. That is:
$U_g U_\Sigma=\tilde{U}_{\Sigma}U_{g}$
with $\tilde{U}_{\Sigma}=U_g U_\Sigma U_g^{-1}$ which remains a product of local operations on the qubits in ${\cal C}_{O}(g)$. We then just apply the {\it decoding operator} $\tilde{U}_{\Sigma}^{\dag}$ to the output cluster qubits in order to wash out the effect of the byproduct operator. On the other hand, if $g$  is not in the Clifford group, we obtain $U_g U_\Sigma=U_\Sigma \tilde{U}_g$, while still keeping the local nature of the byproduct operator. In this case, the measurement pattern must be changed in order to account for the propagation effect~\cite{RBH}.

\section{Concatenation of basic building blocks}
\label{tessellation}

It is clear that the important elements in the \QC~model are the design of the appropriate measurement pattern, along with a complete knowledge of the corresponding decoding operator. Within the stabilizer formalism, the pattern ${\cal M}^{{\cal C}_M(g)}$ suitable for a gate $g$ is engineered by manipulating the eigenvalue equations which characterize the cluster state~\cite{RBH}. For large $N$ and $n$, it is not hard to believe that this procedure becomes rapidly unmanageable and an alternative approach to the construction of exploitable cluster configurations is required. 

This necessity is justified under practical perspectives as well. Indeed, as we discuss in the next sections, the realization of a given cluster state is a system-dependent issue. A method particularly suited for condensed-matter systems (as well as optical lattices,  discussed in Section~\ref{modellonostro}), is the creation of regular cluster configurations from which redundant qubits, not necessary for the computational steps performed and thus superfluous in the corresponding cluster's layout, are removed through appropriate single-qubit measurements. This issue will be reprised in more detail in Section~\ref{computation}. In ideal conditions, the removals are harmless for the performance of a particular gate simulation. But when (intrinsic or external) decoherence is introduced in the cluster, this is no longer true. One of the key results in~\cite{noiarchivio}, is that the number of measurements performed on the cluster qubits (either within ${\cal M}^{{\cal C}_M(g)}$ or instrumental to the cluster state generation) should be as small as possible. Thus, tailoring a cluster state by cleansing it from the redundant qubits may not be an efficient strategy. 

On the other hand, very recent experimental efforts in the all-optical scenario have demonstrated that small clusters of just a few qubits can be efficiently produced within state of the art technology~\cite{vlatko,pan}. In these cases, the setup is built so that {\it exactly} the layout needed is produced and no qubit has to be thrown away. Small optical clusters can then be mutually connected by gluing them as suggested in~\cite{brownrudolph}. Here, we describe an approach to cluster state information processing which uses fundamental configurations or $BBB$'s. By concatenating them, {\it i.e.} placing the blocks next to each other such that they overlap, any desired layout corresponding to a precise computational task can be constructed and the appropriate decoding operator $\tilde{U}^{\dag}_{\Sigma}$ is naturally retrieved. This concatenation technique gives us the possibility of designing optimal elementary steps, each corresponding to one stage of an experiment, which limit the effects of the intrinsic model for imperfection introduced in~\cite{noiarchivio} and discussed in more detail here. 

\subsection{Basic building blocks and concatenation}
\label{bbb}

We start by introducing the $BBB$'s and give their equivalent network circuits. Our diagrammatic notation is such that each qubit will be represented by the symbol of a square. The angle $\alpha$ inside the $i$-th qubit symbol identifies the basis $B_i(\alpha):=\{ \ket{+}^{\alpha}, \ket{-}^{\alpha}\}$ in which that qubit is measured. Here, $\ket{\pm}^{\alpha}=(1/{\sqrt 2})(\ket{0} \pm e^{i \alpha} \ket{1})$ with $\alpha$ the angle from the +ve $x$ axis in the $xy$ plane of the Bloch sphere and $s_i^{xy(\alpha)} \in \{ 0,1\}$ is the corresponding measurement outcome. Empty squares stand for unmeasured (output) qubits. In the configurations we show here, the orientation is irrelevant so long as the topology does not change.

\begin{figure}[ht]
\hspace*{-0.6cm}{\bf (a)}\hspace*{3.0cm}{\bf (b)}\\
\vspace*{0.3cm}
\begin{minipage}[c]{0.1\textwidth}
\psfig{figure=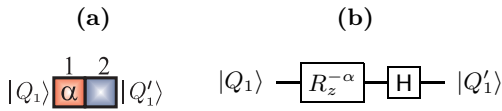,width=2.1cm,height=0.7cm}
\end{minipage}%
\hskip0.8cm
\begin{minipage}[l]{0.2\textwidth}
\mbox{ \Qcircuit @C=1em @R=.7em {
  \lstick{\ket{Q_1}} & \gate{R_z^{-\alpha}} & \gate{\sf H} & 
  \rstick{\ket{Q_1'}} \qw }}\end{minipage}
  \centerline{}
\caption{{\bf (a)}: The $BBB_1$ layout. {\bf (b)}: The operation simulated when qubit $1$ is measured in the $B_1(\alpha)$ basis and $s_1^{xy(\alpha)}=0$ is obtained.}
\label{fig:tfig1}
\end{figure}

\begin{figure}[ht]
{\bf (a)}\hskip3.1cm{\bf (b)}\\
\vspace*{0.3cm}
\begin{minipage}[c]{0.1\textwidth}
\psfig{figure=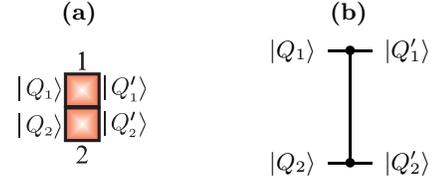,width=1.8cm,height=1.5cm}
\end{minipage}%
\begin{minipage}[l]{0.2\textwidth}
\hskip1.7cm
\mbox{
\Qcircuit @C=1em @R=4.6em {
  \lstick{\ket{Q_1}} & \ctrl{1} &
  \rstick{\ket{Q_1'}} \qw \\
  \lstick{\ket{Q_2}} & \ctrl{-1} &
  \rstick{\ket{Q_2'}} \qw }}
\end{minipage}
\centerline{}
\caption{The $\sf C_{\pi}PHASE$ gate simulated by $BBB_2$ on two logical qubits $\ket{Q_1}$ and $\ket{Q_2}$.}
\label{fig:tfig4}
\end{figure}

The smallest cluster state we can conceive consists of two qubits and can be used to simulate a unitary operation on one logical qubit $\ket{Q_1}$ encoded on a physical cluster qubit (thus $\vert{\cal C}_{I}(g)\vert=1$), as shown in Fig.~\ref{fig:tfig1} {\bf (a)}. This configuration will be denoted from now on as $BBB_1$.
The overall operation simulated by $BBB_1$, when the measurement on qubit $1$ gives $s_1^{xy(\alpha)}=0$ as the outcome, is shown in Fig.~\ref{fig:tfig1} {\bf (b)}, where $R^{-\alpha}_{z}$ represents a rotation around the $z$ axis by an angle $-\alpha$ and $\sf H$ denotes a Hadamard gate. Due to the probabilistic nature of the simulation, it is necessary to apply a decoding operator $\tilde{U}_{\Sigma}^{\dag}(s_1^{xy(\alpha)})=\sigma_x^{s_1^{xy(\alpha)}}$ to qubit $2$. Using the same two-qubit cluster layout, with two encoded logical qubits $\ket{Q_1}$ and $\ket{Q_2}$ ({\it i.e.} here ${\cal C}_{I}(g)\equiv{\cal C}_{O}(g)$, see Fig.~\ref{fig:tfig4} {\bf (a)}), we simply simulate a controlled-$\pi$ phase gate ($\sf C_{\pi}PHASE$)~\cite{nielsenchuang}, as indicated in Fig.~\ref{fig:tfig4} {\bf (b)}. We denote this configuration as $BBB_2$.

\begin{figure}[h]
\hskip-0.3cm\centerline{{\bf (a)}\hskip3.6cm{\bf (b)}}\\
\begin{minipage}[c]{0.1\textwidth}
\centerline{\psfig{figure=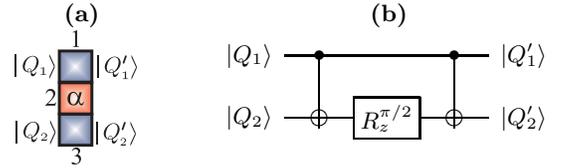,width=1.7cm,height=1.8cm}}
\end{minipage}
\hskip1.8cm
\begin{minipage}[l]{0.2\textwidth}
 \Qcircuit @C=1em @R=1.7em {
  \lstick{\ket{Q_1}}& \ctrl{1}&\qw &\ctrl{1}& 
  \rstick{\ket{Q_1'}}\qw\\
\lstick{\ket{Q_2}}&\targ &\gate{R^{\pi/2}_{z}} & \targ&
  \rstick{\ket{Q_2'}}\qw}
\end{minipage}
\caption{(a): The $BBB_3$ configuration. {\bf (b)}: The equivalent quantum circuit corresponding to the operation simulated by $BBB_3$ with $\alpha=\pi/2$ and $s^{xy(\frac{\pi}{2})}_2=0$.}
\label{fig:ygate1}
\end{figure}

Our set of $BBB$'s is completed by the introduction of $BBB_3$ shown in Fig.~\ref{fig:ygate1} {\bf (a)}, where qubits $1$ and $3$ encode the input states and a measurement is performed on qubit $2$ in the $B_{2}(\alpha)$ basis. This pattern simulates the transformation
\begin{equation}
T_{BBB_3}=\frac{1}{\sqrt 2}
\begin{pmatrix}
1\pm e^{-i\alpha}&0&0&0\\
0&1\mp e^{-i\alpha}&0&0\\
0&0&1\mp e^{-i\alpha}&0\\
0&0&0&1\pm e^{-i\alpha}
\end{pmatrix},
\end{equation}
where the top (bottom) sign corresponds to $s^{xy(\alpha)}_2=0$ ($s^{xy(\alpha)}_2=1$). For $\alpha = \{ 0, \pi \}$ {\it i e.} a measurement of the bridging qubit in the $\pm \sigma_x$ eigenbasis, one obtains a non-unitary gate. We are particularly interested in what is found for $\alpha=\frac{\pi}{2}$, which corresponds to an operation decomposed as shown in Fig.~\ref{fig:ygate1} {\bf (b)}. The decoding operator for this specific gate simulation is $\tilde{U}_{\Sigma}^{\dag}=-i\sigma^{(1)s^{xy(\pi/2)}_2}_{z}\otimes\sigma^{(3)s^{xy(\pi/2)}_2}_{z}$. 

\begin{figure}[ht]
\begin{minipage}[c]{0.1\textwidth}
\psfig{figure=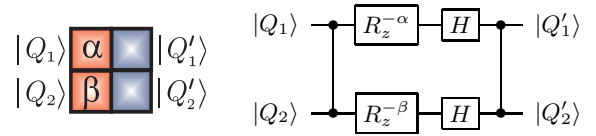,width=2.6cm,height=1.2cm}
\end{minipage}%
\hskip0.5cm
\begin{minipage}[l]{0.25\textwidth}
\mbox{
\Qcircuit @C=1em @R=2.0em {
  \lstick{\ket{Q_1}} & \ctrl{1} & \gate{R_z^{-\alpha}} & \gate{H} & \ctrl{1} &
  \rstick{\ket{Q_1'}} \qw \\
  \lstick{\ket{Q_2}} & \ctrl{-1} & \gate{R_z^{-\beta}} & \gate{H} & \ctrl{-1} &
  \rstick{\ket{Q_2'}} \qw }}
\end{minipage}
\centerline{}
\caption{The concatenation of two $BBB_1$'s (Fig.~\ref{fig:tfig1}) with two $BBB_2$'s (Fig.~\ref{fig:tfig4}).}
\label{fig:tfig7}
\end{figure}

\begin{figure}[ht]
\begin{minipage}[c]{0.1\textwidth}
\psfig{figure=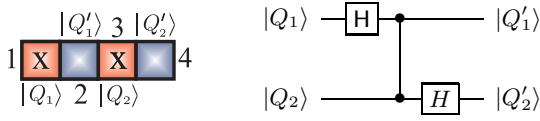,width=2.5cm,height=1.4cm}
\end{minipage}%
\hskip0.5cm
\begin{minipage}[l]{0.25\textwidth}
\mbox{
\Qcircuit @C=1em @R=2.0em {
  \lstick{\ket{Q_1}} & \gate{\sf H} & \ctrl{1} & \qw& \rstick{\ket{Q_1'}} \qw \\
  \lstick{\ket{Q_2}} & \qw   & \ctrl{-1} & \gate{H}& \rstick{\ket{Q_2'}} \qw }}
\end{minipage}
\centerline{}
\caption{The concatenation of two $BBB_1$'s and one $BBB_{2}$ with measurements in the $B_1(0)$ and $B_3(0)$ basis.}
\label{fig:tfig8}
\end{figure}

\begin{figure}[ht]
\hskip-1.7cm{\bf (a)}\hskip3.9cm{\bf (b)}\\
\vspace*{0.2cm}
\hskip-1.2cm
\begin{minipage}[c]{0.1\textwidth}
\psfig{figure=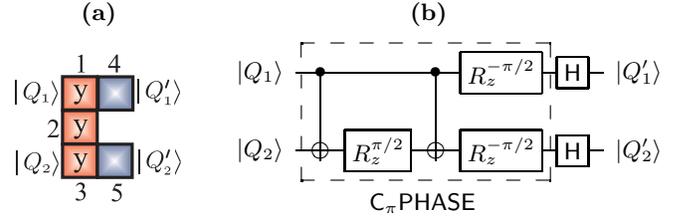,width=2.3cm,height=2.0cm}
\end{minipage}%
\hskip1.4cm
\begin{minipage}[l]{0.2\textwidth}
\mbox{
 \Qcircuit @C=0.6em @R=1.5em {
  \lstick{\ket{Q_1}}& \ctrl{1}&\qw &\ctrl{1}& \gate{R^{-\pi/2}_{z}}&\gate{\sf H}& \rstick{\ket{Q_1'}}\qw\\
\lstick{\ket{Q_2}}&\targ &\gate{R^{\pi/2}_{z}} & \targ&\gate{R^{-\pi/2}_{z}}&\gate{\sf H}& \rstick{\ket{Q_2'}}\qw\gategroup{1}{2}{2}{5}{.7em}{--}}}
\centerline{}
\hskip0.8cm{$\sf C_{\pi}PHASE$}
\end{minipage}\\
\caption{{\bf (a)}: The configuration obtained by concatenating two $BBB_1$'s with a $BBB_3$. The single-qubit measurements are in the $B_{1\rightarrow{3}}(\pi/2)$ bases.  {\bf (b)}: The corresponding equivalent quantum circuit. The boxed part is equivalent to a $\sf C_{\pi}PHASE$.}
\label{fig:ygate}
\end{figure}

Using this set of $BBB$'s, we can construct more complicated configurations and in order to see this more clearly, we use the following observation from~\cite{RBH}:

\begin{Theorem}
\label{one}
Consider a quantum circuit $g$ associated with the unitary
operator $U_g$ simulated on a cluster ${\cal C}(g)$. Let $g$ 
be comprised of two consecutive circuits $g_1$ and $g_2$ on
subclusters ${\cal C}(g_1)$ and ${\cal C}(g_2)$ respectively, {\it i.e.} 
$g=g_2 g_1$ and ${\cal C}(g)={\cal C}(g_1)\cup{\cal C}(g_2)$, with
${\cal C}(g_1)\cap{\cal C}(g_2)$ containing one cluster qubit
for each logical qubit. These subgates have associated unitary
operators $U_{g_1}$ and $U_{g_2}$. The method of entangling the
whole cluster ${\cal C}(g)$ and performing the required measurements
for simulating $g$, is equivalent to entangling the qubits of
${\cal C}(g_1)$, performing the required measurements for $g_1$,
then entangling the qubits of ${\cal C}(g_2)$ and performing the
required measurements for $g_2$.
\end{Theorem}
\begin{proof}
In the first method, the logical input state $\ket{\psi_{\rm in}}$
is encoded onto the cluster qubits in the input section ${\cal
C}_I(g_1)$ of the first subcluster. The entire cluster ${\cal
C}(g)$ is then entangled and qubits in ${\cal C}(g)\setminus{\cal
C}_O(g_2)$ are measured in the required bases, with the logical
output state $\ket{\psi_{\rm out}}$ present on the cluster qubits
of ${\cal C}_O(g_2)$.

In the second method, the logical input state $\ket{\psi_{\rm in}}$
is again encoded onto the cluster qubits in the input section ${\cal
C}_I(g_1)$ of the first subcluster. The subcluster ${\cal C}(g_1)$
is then entangled and qubits in ${\cal C}(g_1)\setminus{\cal
C}_O(g_1)$ are measured in the required bases, with the
intermediate logical output state $\ket{\psi_{\rm out}'}$ present
on the cluster qubits of ${\cal C}_O(g_1)={\cal C}_I(g_2)$. A
similar procedure for the second subcircuit $g_2$ is carried out
and the logical output state $\ket{\psi_{\rm out}''}$ is present
on the cluster qubits of ${\cal C}_O(g_2)$.

To show the two methods are equivalent, is to show that
$\ket{\psi_{\rm out}} = \ket{\psi_{\rm out}''}$. Let $P_1,~P_2$ be
projectors representing the measurements on qubits in ${\cal
C}(g_1)\setminus{\cal C}_O(g_1)$ and ${\cal C}(g_2)\setminus{\cal
C}_O(g_2)$ respectively with $S^{({\cal C}(g_1))}=:S_1$ and
$S^{({\cal C}(g_2))}=:S_2$ representing the entanglement
operations on subclusters ${\cal C}(g_1)$ and ${\cal C}(g_2)$
respectively. As $P_1$ commutes with $S_2$, we find that
\begin{equation}\label{eq:theorem1}
P_2S_2P_1S_1=P_2P_1S_2S_1,
\end{equation}
This mixture of entangling operators and projectors acts on the state $\ket{\psi_{\rm in}}_{{\cal
C}_I(g)} \otimes_{a \in {\cal C}(g)\setminus{\cal
C}_I(g)} \ket{+}_a$. As they act equivalently, we have that $\ket{\psi_{\rm out}} = \ket{\psi_{\rm out}''}=U_gU_{\Sigma}\ket{\psi_{\rm in}}$ and the two methods are equivalent.
\end{proof}
Observation~\ref{one} can be extended to an arbitrary number of sub-gates
$g_i$ associated with unitary operators $U_{g_i}$ on sub-clusters
${\cal C}(g_i)$. These then make up the gate $g$ associated with
the operator $U_g$ on the entire cluster ${\cal C}(g)$. 
For each projector $P_i$ there is an associated unitary operator $U_{g_i}$ applied to the logical qubits. However $U_{g_i}$ is not the only operation applied. Due to the fact that there are two possible outcomes for each qubit measured, the gate simulation is probabilistic. It turns out that for all the $BBB$'s in the last section, the actual unitary operation implemented can be written as $U_{g_i}U_{\Sigma}(s_i)$ ($g_i=BBB_i,\,i=1,2,3$), where $U_{\Sigma}(s_i)$ is a local byproduct operator
dependent on the outcome $s_i:=s^{xy(\alpha_j)}_j$ of the measurement performed on the $j$-th qubit in the ${\cal C}(g_i)\setminus{\cal C}_O(g_i)$ part of the
cluster.
In this way the unitary operation $U_{g_i}$ becomes deterministic, as we can
always propagate $U_{\Sigma}(s_i)$ through it, keeping
the byproduct operator local. For a large concatenation of
subclusters, this propagation procedure can be carried out for
$U_{g_i}$ and $U_{\Sigma}(s_a,s_b,s_c)=\sigma^{s_a}_x\sigma^{s_b}_y\sigma^{s_c}_z$, where $s_a,s_b,s_c \in \{0,1 \}$ are a result of the byproduct operators corresponding to the gates acting before $U_{g_i}$. However, when $U_{g_i}$ is not in the Clifford group, it will change upon propagation of the byproduct operators through it. For all of the $BBB$'s mentioned before, we find that this only occurs with $R_z^{-\alpha} \sigma_x^{s_a}$ and $R_z^{-\alpha} \sigma_y^{s_b}$. We may write
\begin{eqnarray}
R_z^{-\alpha} \sigma_x^{s_a}=\sigma_x^{s_a}R_z^{-\alpha'}, \nonumber \\
R_z^{-\alpha} \sigma_y^{s_b}=\sigma_y^{s_b}R_z^{-\alpha''},
\end{eqnarray}
where $\alpha'=(-1)^{s_{a}}\alpha$ and $\alpha''=(-1)^{s_{b}}\alpha$. In other words the measurements become adaptive, but the resources (in terms of the number of cluster qubits) never
change, the byproduct operators always remain local and we can counteract the effect of the byproduct operators by changing the measurement angle from $\alpha$ to $-\alpha$.
Overall in a large concatenated circuit of $BBB$'s, we end up
with the logical input $|\psi_{\text{in}}\rangle$ and output
$|\psi_{\text{out}}\rangle$ of the unitary simulation
$U_g=\prod_{i=|{\cal{N}}|}^{1} \, U_{g_i}$, related via
\begin{equation}
    \label{Gateseq}
    |\psi_{\text{out}}\rangle = \left( \prod_{i=|{\cal{N}}|}^{1}
    \tilde{U}_{\Sigma}(s_i)\right) \, \left( \prod_{i=|{\cal{N}}|}^{1}\tilde{U}_{g_i}\right)
     |\psi_{\text{in}}\rangle.
\end{equation}
Here, ${\cal{N}}$ is the number of subgates, and the
$~\tilde{\phantom{}}~$ represents the fact that the byproduct
operators change on propagation, or the unitary operations of the
subclusters become adaptive. 
A final note to add is that some byproduct operators will have to propagate {\it further} than
others in this scheme. By applying the hermitian conjugate $\tilde{U}_{\Sigma}^{\dag}(s_a,s_b,s_c)=( \prod_{i=|{\cal{N}}|}^{1}
    \tilde{U}_{\Sigma}(s_i))^{\dag}$ of the
propagated byproduct operators we will recover the unitary operation desired. The $\tilde{U}_{\Sigma}^{\dag}(s_a,s_b,s_c)$ are decoding operators and for each logical qubit, they will always be a multiple from the set $\{\openone,\sigma_x,\sigma_z,\sigma_y\}$ up to a global phase factor $\in \{1,-1 \}$. We therefore recover the cluster state model for QC, without the need for the stabilizer formalism and eigenvalue equations of Eq. (\ref{gatecheck}). It is now easier to design the cluster configuration which simulates a desired quantum circuit and we are on our way 
toward the construction of computationally useful {\it extended} building blocks ($EBB$'s). By concatenating two $BBB_1$'s and two $BBB_2$'s, we produce a bidimensional square cluster state, the {\it box cluster}, which has been experimentally realized and used in order to perform a two-qubit quantum searching algorithm~\cite{vlatko}. The physical layout, the measurement pattern and the corresponding equivalent circuit are shown in Fig.~\ref{fig:tfig7}.
On the other hand, by concatenating two $BBB_1$'s and a $BBB_2$, we obtain a simple four-qubit linear cluster which is particularly interesting. Indeed, if the single-qubit measurements are performed in the $\sigma_x$-eigenbasis, the corresponding equivalent quantum circuit is locally equivalent to a ${\sf CNOT}$ gate {\it i.e.} the read-out bases of qubits $2$ and $4$ are the $\sigma_x$ eigenbases~\cite{noiarchivio}. The layout and the measurement pattern are shown in Fig.~\ref{fig:tfig8}.
We will comment on the importance of this cluster configuration in Section~\ref{computation}, where its role in an economical scheme for cluster state based QC is highlighted~\cite{noiarchivio}. For our purposes here, it is instructive to explicitly compute the byproduct operator corresponding to this gate simulation. 
As we have said, the decoding operator needed after the measurement pattern in $BBB_1$ is given by either $\openone$ or $\sigma_x$, depending on the outcome of the measurement performed. Thus, including the random action of the byproduct operators, the transformation in Fig.~\ref{fig:tfig8} can be represented overall by
\begin{equation}
\label{byp}
\left(\openone^{(2)}\otimes\sigma^{(4)s^{xy(0)}_{3}}_x\right){\sf H}_4{\sf C_{\pi}PHASE}\left(\sigma^{(2)s^{xy(0)}_{1}}_x\otimes\openone^{(4)}\right){\sf H}_2.
\end{equation}
Note that no rotation operator appears in the above expression because of the choice of $B_{1}(0)$ and $B_3(0)$ for the measurement pattern. We now use the fact that ${\sf C_{\pi}PHASE}(\sigma^{(i)}_x\otimes\openone^{(j)})=(\sigma^{(i)}_x\otimes\sigma^{(j)}_z){\sf C_{\pi}PHASE}$ and that ${\sf H}_i\sigma^{(i)}_z=\sigma^{(i)}_x{\sf H}_i$, so that the $(\sigma^{(2)s^{xy(0)}_{1}}_x\otimes\openone^{(4)})$ part in Eq.~(\ref{byp}) can be propagated until we end up with $\tilde{U}_\Sigma{\sf H}_4S^{24}{\sf H}_{2}$, where
\begin{equation}
\label{byp2}
\tilde{U}_{\Sigma}=\sigma^{(2)s^{xy(0)}_{1}}_x\otimes\sigma^{(4)[s^{xy(0)}_{1}\oplus{s}^{xy(0)}_{3}]}_x,
\end{equation} 
with $\oplus$ as the logical XOR operation in this case only.
This result, originally presented in~\cite{noiarchivio}, is now rigorously demonstrated by a simple argument based on the concatenation technique.
Our $EBB$ set is completed by the configuration in Fig.~\ref{fig:ygate} {\bf (a)}, {\it i.e.} the concatenation of one $BBB_3$ and two $BBB_1$'s. The bases for the measurement pattern are $B_{1\rightarrow{3}}(\alpha=\pi/2)$ and qubits $1$ and $3$ encode the input states. The overall transformation is decomposed as shown in Fig.~\ref{fig:ygate} {\bf (b)}.
Following the lines depicted above, the propagated byproduct operator corresponding to this gate simulation is $\tilde{U}_\Sigma=\sigma^{s^{xy(\pi/2)}_1{\oplus}s^{xy(\pi/2)}_2}_x\otimes\sigma^{s^{xy(\pi/2)}_3{\oplus}s^{xy(\pi/2)}_2}_x$. We will see in Section~\ref{computation} that this $EBB$ plays a central role in the original {\sf CNOT} simulation discussed in~\cite{RBH}. 

It is remarkable that the explicit form of the byproduct operators has been found by using the expressions valid for the $BBB$'s, the concatenation rule and basic commutation properties. The stabilizer formalism is not required anymore and the advantages of this alternative technique become clear once complicated cluster configurations, involving many qubits, need to be considered. We thus drop the stabilizer formalism and adopt this technique for the remainder of the paper.
\section{Imperfect Generation of a Cluster State} 
\label{modellonostro}

The key element in generating a cluster state is the ability to perform the controlled gates $S^{ab}$ on the qubits occupying the sites of a cluster ${\cal C}$. However, these operations can be inherently imperfect or not precisely controllable. For example, an optical lattice loaded by a bunch of neutral atoms is a candidate for the embodiment of a \QC~\cite{bloch,mouraalves,garciaripoll}. In this particular setup, any two-qubit interaction is realized through controlled collisions of the atoms loading the optical lattice. Several experiments~\cite{bloch} consider a unidimensional lattice loaded by bosonic atoms with two hyperfine levels $\ket{c_{1}}$ and $\ket{c_{2}}$. The state of an atom occupying the $j$-th site in the lattice, prepared in its internal state $\ket{c_{1}}$ ($\ket{c_{2}}$), can be described by the application of the bosonic creation operator $\hat{c}^{(j)\dag}_{1}$ ($\hat{c}^{(j)\dag}_{2}$) to some fiducial initial atomic state. The Hamiltonian describing the atomic interaction takes the form of a two-species Bose-Hubbard model~\cite{garciaripoll}
\begin{eqnarray}\label{mandel}
H_{ol}&=&\frac{U}{2}\sum^{N}_{j=1}(\hat{c}^{(j)\dag}_{1}\hat{c}^{(j)\dag}_{1}\hat{c}^{(j)}_{1}\hat{c}^{(j)}_{1}+\hat{c}^{(j)\dag}_{2}\hat{c}^{(j)\dag}_{2}\hat{c}^{(j)}_{2}\hat{c}^{(j)}_{2} \nonumber \\
&&+2\hat{c}^{(j)\dag}_{1}\hat{c}^{(j)}_{1}\hat{c}^{(j)\dag}_{2}\hat{c}^{(j)}_{2}),
\end{eqnarray}
where $U$ is the coupling strength~\cite{specificaHubbard}. The first two terms in the Hamiltonian describe {\it in-site} free dynamics of the two different {\it species}, while the last one takes into account the interactions responsible for the controlled phases introduced in the joint state of two atoms which occupy the same site. The coupling strength is in general a function of the position in the lattice ({\it i.e.} $U$ has a spatial profile along the linear lattice) and its value can fluctuate due to instabilities in the intensities of the lasers used in order to create the lattice. A fluctuating $U$ gives rise to a random phase shift imposed after the qubit-qubit interaction, which may be different from the $\pi$ value required to attain a perfect cluster state. 
Thus, the efficiency of the $S^{ab}$'s critically depends on the control we have over the strength of the interactions. The initial filling-fraction of the lattice may also influence the performances of the gates and the extent to which entanglement can be spread across the cluster~\cite{bloch,garciaripoll}. If the degree of control is not optimal, we face the problem of imperfect (inhomogeneous) interactions throughout the physical lattice. 
Here, this issue is addressed by formally considering imperfect entangling operations $S^{ab}_{D}$'s on the qubits of the cluster. Our model consists of controlled gates having the form 
\begin{equation}
    \label{Sabdefdirty}
    \Sab_D= |0 \rangle_a \langle 0| \otimes {\openone}^{(b)} + |1 \rangle_a \langle 1| \otimes \left(|0 \rangle_b \langle 0|-e^{i \theta_{a}}|1 \rangle_b \langle
    1| \right),
\end{equation}
which add the phase $\theta_{a}$ to the desired and optimal $\pi$. As in the ideal case, all $\Sab_D$'s mutually commute and the imperfect entangling operation $S^{{\cal{C}}}_D=\prod_{<a,b>}\Sab_D$ is unitary. 

Let us consider the use of Eq.~(\ref{Sabdefdirty}) in order to generate a unidimensional noisy cluster state (a {\it noisy linear cluster state}). By applying
$S^{{\cal{C}}}_D$ to the product state $|+\rangle_{{\cal{C}}}$ of $N$ qubits in a linear cluster, we find~\cite{noiarchivio}
\begin{equation}
    \label{noisyclust}
|\phi \rangle_{{\cal{C}}}^D=S^{{\cal{C}}}_D
|+\rangle_{{\cal{C}}}={2^{-{N}/{2}}}\sum_{{\bf z}_i}\prod^{N-1}_{j=1}(-e^{i\theta_j})^{z^{i}_jz^{i}_{j+1}}|{\bf z}_i\rangle,
\end{equation}
where $z^{i}_j$ is the value of the $j$-th binary digit of the integer ${\bf z}_i$ and the summation runs over all the ${\bf z}_i$ for $i\in[0,2^N-1]$. The properties of the state $|\phi \rangle_{{\cal{C}}}^D$ have been discussed in detail in~\cite{noiarchivio}, where the fidelity and entanglement structure of noisy cluster states are analyzed.  Various probability distributions were placed on the unwanted phases $\theta_j$ in order to give a better idea of how the fidelity might be affected within realistic situations. However it was noted that no universal model for the probability distribution existed. As a result, we provided a more general picture to which specific distributions could then be investigated, such as the Gaussian distribution. It was found that the fidelity of unidimensional and bidimensional clusters rapidly decreases with a rise in the number of qubits, even when small amounts of noise are considered. The entanglement structure was also found to be profoundly affected by the randomness introduced to the unwanted phases. It was shown that the amount of genuine multipartite entanglement becomes reduced to the benefit of bipartite quantum correlations, fragile against fluctuations in the unwanted phases $\theta_j$ and eventually leads to an entanglement breaking effect. From this previous investigation, one can clearly see that QIP protocols should be strongly affected by imperfections in the initial entanglement created between the qubits in a cluster~\cite{noiarchivio}. Here we intend to give a broader picture of the effect of this noise model on various other key QIP protocols.


\section{Information flow across a noisy linear cluster}
\label{informationflow}

We now start our analysis of the performances of one-way protocols for QIP with {\it noisy cluster states}. While Section~\ref{computation} will be dedicated to the basic computational steps, here we deal with a communication issue represented by the transfer of quantum information through a cluster state. We approach the reliability of state transmission along a quantum channel represented by a linear cluster. Quantum state transfer has recently received considerable attention in the context of limited-resource QIP~\cite{qst}. Here, our approach will be using a quantum resource provided by the multipartite entanglement in a cluster state and conditional dynamics given by measurements.

The idea behind the realization of information flow through a cluster state is simple. In the one-way model, seen from the original stabilizer viewpoint, any set of single-qubit measurements performed onto the qubits belonging to ${\cal C}_M(g)$, together with the measurements in the $\sigma_x$-eigenbasis of the qubits in ${\cal C}_I(g)$ (see Section \ref{operation}), processes the encoded input state and at the same time, transfers it to the ${\cal C}_O(g)$ section. Thus, one can naively think about simulating the identity gate $g=\openone$ using an appropriate cluster state, where the transfer of the input state to the output section of the cluster occurs naturally. However it is much simpler to think of this information flow in a linear cluster in terms of the concatenation of many $BBB_1$'s. In this case, measuring qubit $1$ in the $B_1(0)$ basis, means the encoded state is transferred to $2$, while at the same time rotated by $\sf H$. Thus, forgetting for the moment the effect of the byproduct operator, by arranging the concatenation of many $BBB_1$'s, {\it i.e.} a pattern of measurements in the single-qubit $\sigma_x$-eigenbasis for all the qubits in ${\cal C}_I(g)\cup{\cal C}_M(g)$ of a sufficiently long linear cluster, either we realize the information flow we are looking for (when the number of effective Hadamard gates is even) or we simply get $\ket{\psi_{\rm out}}={\sf H}\ket{\psi_{\rm in}}$ which can be corrected once we know the parity of $N$. 

The question we raise here is about the efficiency of the transfer process when noisy cluster states are used. Ideally the fidelity for information transfer would be equal to unity. However, this is not the case if noisy linear clusters are used. In order to obtain a full picture of the protocol for information flow, we need to calculate the form of the proper decoding operator arising after the concatenation of many $BBB_1$'s. 

\subsection{Identity and Hadamard gate}
\label{identita&hadamard}

We already know that $BBB_1$ with $B_1(0)$ effectively simulates {\sf H} on a logical qubit $\ket{\psi_{\rm in}}$, with $\tilde{U}_\Sigma=\sigma^{s^{xy(0)}_1}_x$. If we now concatenate two $BBB_1$'s to obtain a three-qubit linear cluster, by measuring qubits $1$ and $2$ we simulate 
\begin{equation}
\sigma^{s^{xy(0)}_{2}}_x{\sf H}\sigma^{s^{xy(0)}_{1}}_x{\sf H}=\underbrace{\sigma^{s^{xy(0)}_{2}}_x\sigma^{s^{xy(0)}_{1}}_z}_{\tilde{U}_\Sigma}\openone
\end{equation} 
on the logical qubit $\ket{\psi_{\rm in}}$. The decoding operator to decode the information transferred to and stored on qubit $3$ is then given by $\tilde{U}_{\Sigma}^{\dag}=\sigma^{s^{xy(0)}_{1}}_z\sigma^{s^{xy(0)}_{2}}_x$. The advantage of using this concatenation technique with respect to the stabilizer formalism is evident and even more striking when larger linear clusters are considered. 
\subsection{Flowing information}
\label{trasferimento}

\begin{figure}
\includegraphics[width=2.5in]{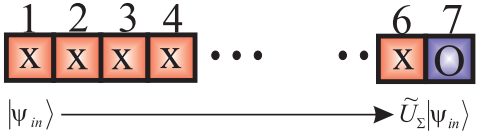}
\caption{\label{fig:fig7} Information flow through an N-qubit linear cluster.}
\end{figure}

In the concatenation model, for an arbitrarily long odd-$N$ qubit quantum channel, the concatenation of $N-1$ $BBB_1$'s results in $\prod^{1}_{i=(N-1)}(\sigma^{s^{x}_{i}}_x{\sf H})$ applied to the input state $\ket{\psi_{\rm in}}$, where $s^{x}_i=s^{xy(0)}_i$. Each time the $x$-Pauli matrix corresponding to an even-labelled qubit is propagated through an {\sf H}, thus being transformed into a $z$-Pauli matrix, the product ${\sf HH}=\openone$ is obtained and the $x$-Pauli matrices of odd-labelled qubits remain unchanged. This results in all the even labels (odd labels) becoming associated with $\sigma_z$ ($\sigma_x$) and we obtain the following structure for the propagated byproduct operator
\begin{equation}
\label{byproductidentity}
\tilde{U}_\Sigma=\sigma_z^{s^x_1 \oplus
(\sum_{i=2}^{(N-1)/2}s^x_{2i-1}){\rm mod} 2} \sigma_x^{s^x_2
\oplus (\sum_{i=2}^{(N-1)/2}s^x_{2i}){\rm mod} 2}~.
\end{equation}

So far however, we have just dealt with ideal cluster states. We now apply the above protocol for the transfer of information encoded in $\ket{\psi_{\rm in}}= a |0 \rangle + \sqrt{1-a^2} |1 \rangle$ across an $N$-qubit noisy linear cluster. The theoretical analysis we have developed so far can still be applied to the case of $S^{ab}\rightarrow S^{ab}_D$ {\it i.e.} the form of the byproduct operator is the same as Eq.~(\ref{byproductidentity}). We have simulated the measurement pattern for information flow on an encoded noisy linear cluster state. Despite the computational challenge represented by the exponentially growing number of different outcome sets (for an $N$-qubit linear cluster, there are $2^{N}$ different sets $\{s^{x}_{j}\},\,j\in[1,N]$), it has been possible to explicitly calculate the state transfer fidelity ${\cal F}_N(a,\theta_j)=|\langle{\psi_{\rm in}}|\psi_{out}\rangle|^2$ for each outcome set, up to $N=9$ qubits. This enables us to evaluate $\tilde{\cal F}_N(a,\theta_j)$, {\it i.e.} the state fidelity {\it averaged} over all the different sets of measurement outcomes, after the application of the relevant decoding operator given by the hermitian conjugate of Eq.~(\ref{byproductidentity}). This quantity gives us an estimate of the average performance of the transfer process. It is important to note that this approach is only one of the possibilities available, the other being the postselection of the event corresponding to a favorable outcome configuration. For example, one could choose to discard all the events but the one where $s^{x}_{j}=0\,\forall{j}\in[1,N]$ as this case corresponds to $\tilde{U}_{\Sigma}^{\dag}=\openone^{(N)}$ and therefore no local adjustment is required after the simulation of the identity gate. 

\begin{figure}[t]
{\bf (a)}\hskip4cm{\bf (b)}
\centerline{\psfig{figure=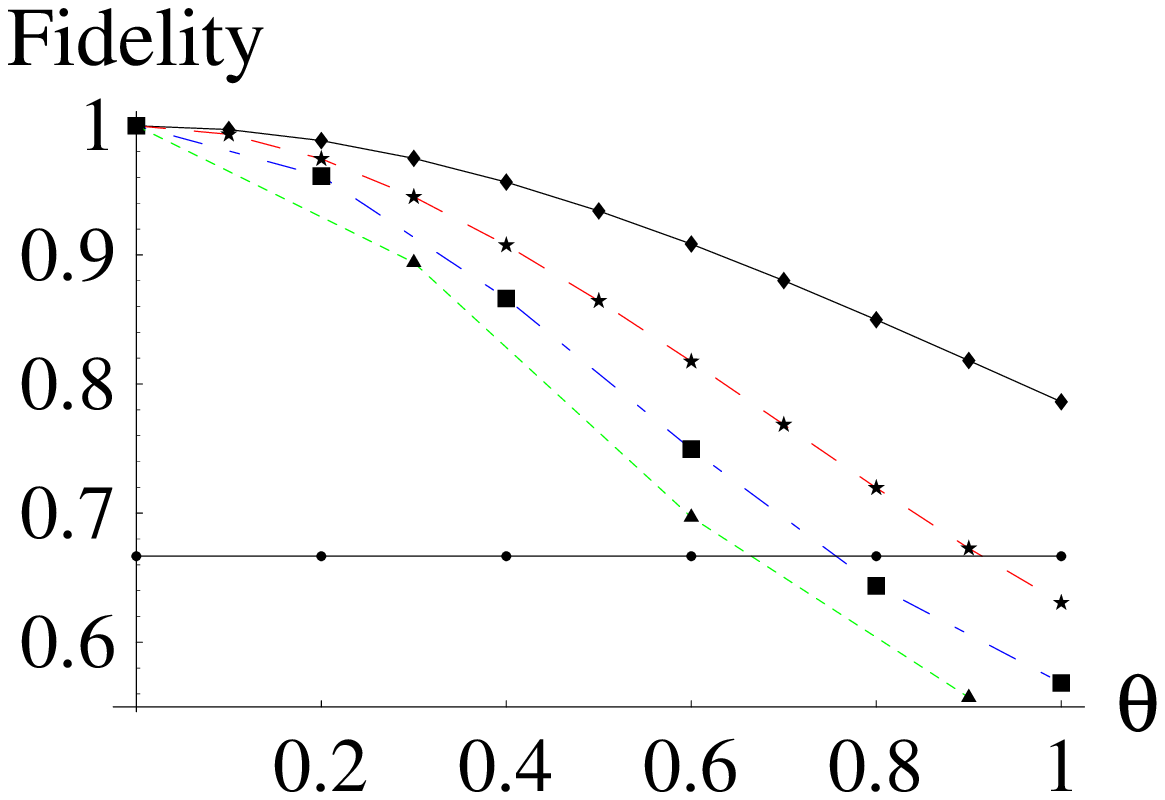,width=4.5cm,height=3.0cm}\psfig{figure=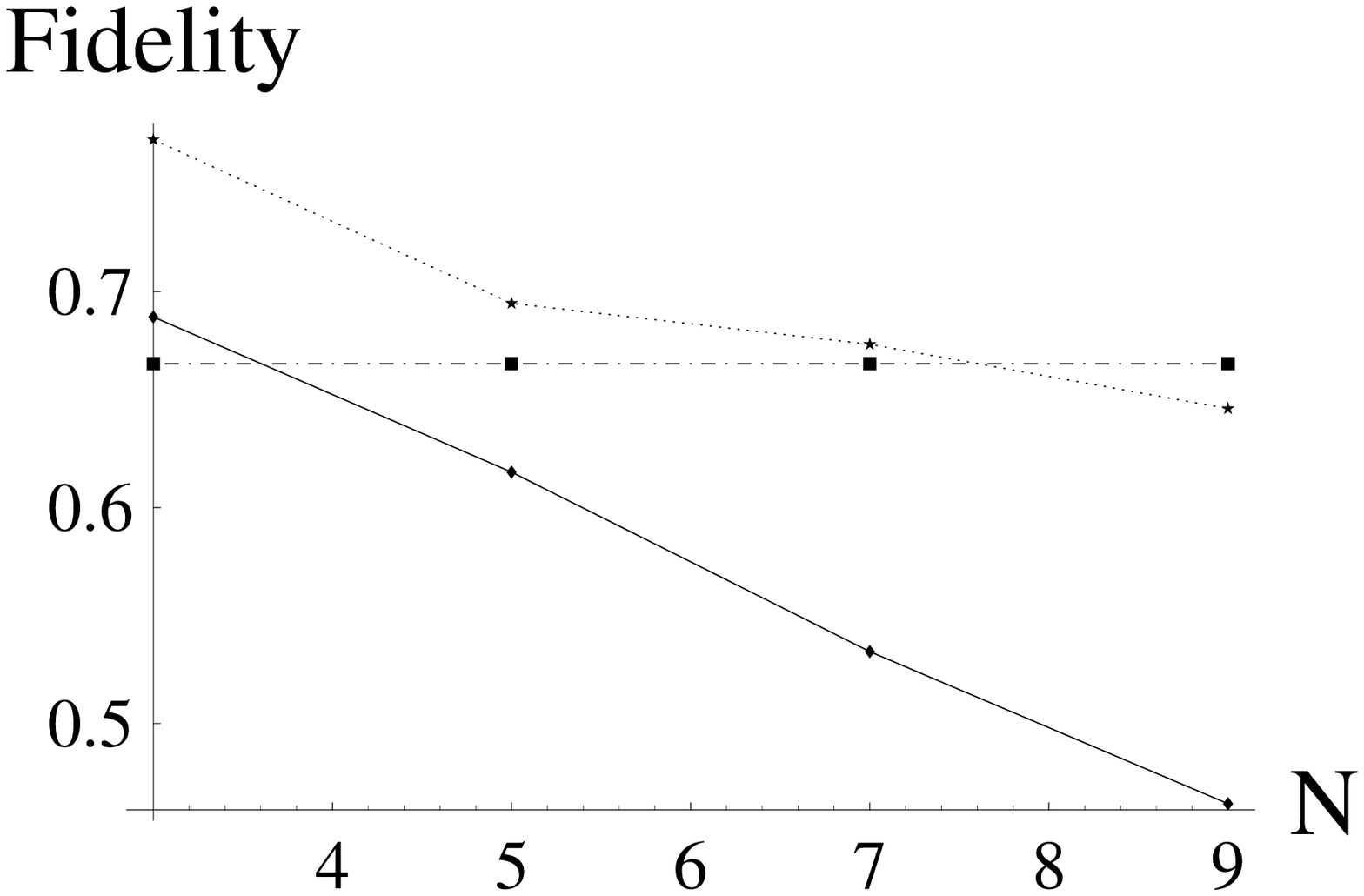,width=4.5cm,height=3.0cm}}
\caption{
{\bf (a)}: Fidelity of information transfer with $a$ averaged over the single-qubit Bloch sphere and all phases $\theta_j =\theta~(\forall{j})$. From top to bottom curve, we show $N=3$ ($\blacklozenge$), $N=5$ ($\bigstar$), $N=7$ ($\blacksquare$) and $N=9$ ($\blacktriangle$). The horizontal line ($\bullet$) represents the classical threshold $2/3$. {\bf(b)}: Fidelity of information transfer with $a$ averaged over the Bloch sphere and $\theta_j$ averaged over Gaussian distributions, centered on $\theta_j=0$, with standard deviation $\sigma=0.5$ ($\bigstar$, dotted line) and $\sigma=1$ ($\blacklozenge$, solid line). Again, the classical threshold is shown for comparison ($\blacksquare$).} 
\label{fig:fig9}
\end{figure}

However, in order to perform a more quantitative investigation and wash out any initial state-dependence, the average state transfer fidelity for any input state $\ket{\psi_{\rm in}}$ must be considered~\cite{qst}. This can be done by assuming a uniform distribution for $a$ and integrating over the Bloch sphere surface as $\bar{\cal F}_N(\theta)=(1/4\pi)\int\tilde{\cal F}_N(a,\theta){d}\Sigma$, where $d\Sigma$ is the surface element. The comparison between $\tilde{\cal F}_{N}(a,\theta_{j})$ and $\bar{\cal F}_N(\theta)$ reveals an almost uniform behavior of ${\cal F}_N(a,\theta)$ with $a$. On the other hand, the integration over the Bloch sphere's surface allows us to compare $\bar{\cal F}_N(\theta)$ with $2/3$, the best fidelity achievable by measuring an unknown qubit state along a random direction and sending the result through a classical channel~\cite{horodecki} (see also Bose and Paternostro {\it et al.}~\cite{qst}). The analysis of $N=3,5,7$ and $9$-qubit cases in Fig.~\ref{fig:fig9} {\bf (a)} reveals that as soon as $\theta\simeq{0.65}$, $\bar{\cal F}_9(0.65)<2/3$. By increasing $\theta$, all the other transfer fidelities (except the case of $N=3$) become worse than the classical threshold value and are thus useless for quantum state transfer. The range of exploitability of the cluster channel shown in Fig.~\ref{fig:fig9} {\bf (a)} is considerably small if probability distributions attached to the $\theta_j$'s are considered. We have quantitatively addressed the case of individual Gaussians, centered on $\theta_j=0$ and with increasing standard deviation $\sigma$. In Fig.~\ref{fig:fig9} {\bf (b)}, we show the totally averaged fidelity ${\cal F}_{N,\sigma}\propto\int{p}(\theta,\sigma)\bar{\cal F}_{N}(\theta)d\theta$, for a Gaussian $p(\theta,\sigma)$ which retains a parameterisation in $N$ and the standard deviation $\sigma$. For $\sigma=1$, it can be seen that already for $N\ge{5}$ the cluster channel becomes less efficient than the best classical strategy for transfer. On the other hand, by reducing the amount of randomness in the noisy cluster state {\it i.e.} by reducing the spread of the distribution down to $\sigma=0.5$, the usefulness of the transfer protocol is restored but only for cluster channels of just a few qubits ($N<{9}$). Thus, the study of state transfer through linear cluster states coincides with the conclusions we have drawn about the state fidelity in~\cite{noiarchivio}: whenever intrinsic random imperfections are considered within the cluster state model, the dimension of the register plays a critical role in the efficiency of the QIP protocols.

\section{Computation with noisy cluster states}
\label{computation}

We are now in a position to investigate the performance of \QC~within the framework of noisy cluster states. In this Section, we address both single-qubit rotations and two-qubit entangling gates, paying particular attention to the {\it paradigm} of this class of gates, the controlled-{\sf NOT} ({\sf CNOT}). The gate fidelity relative to these operations will be thoroughly studied against the effects of the unwanted phases introduced by $S^{ab}_{D}$'s. 
\subsection{Arbitrary rotation}
\label{rotazioni}

\begin{figure}[b]
{\bf (a)}\hspace*{4.3cm}{\bf (b)}\\
\vskip0.4cm
\centerline{\psfig{figure=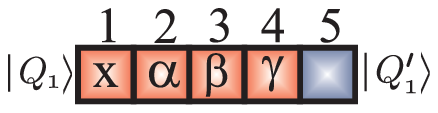,width=4.2cm,height=1.0cm}\hspace*{0.3cm}\psfig{figure=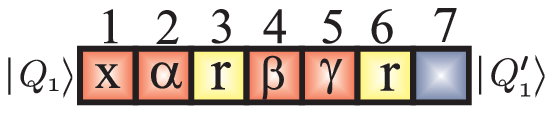,width=4.7cm,height=1.0cm}}
\caption{{\bf (a)}: The layout for a single-qubit rotation where the input logical state encoded on qubit $1$, is rotated by $U_R$ and transfered to qubit $5$ after the measurements shown. {\bf (b)}: A modified configuration with two redundant qubits (qubit $3$ and $6$) which are removed via $\sigma_x$ measurements.}
\label{genrot}
\end{figure}

\begin{figure}[b]
{\bf (a)}\hspace*{4cm}{\bf (b)}
\centerline{\psfig{figure=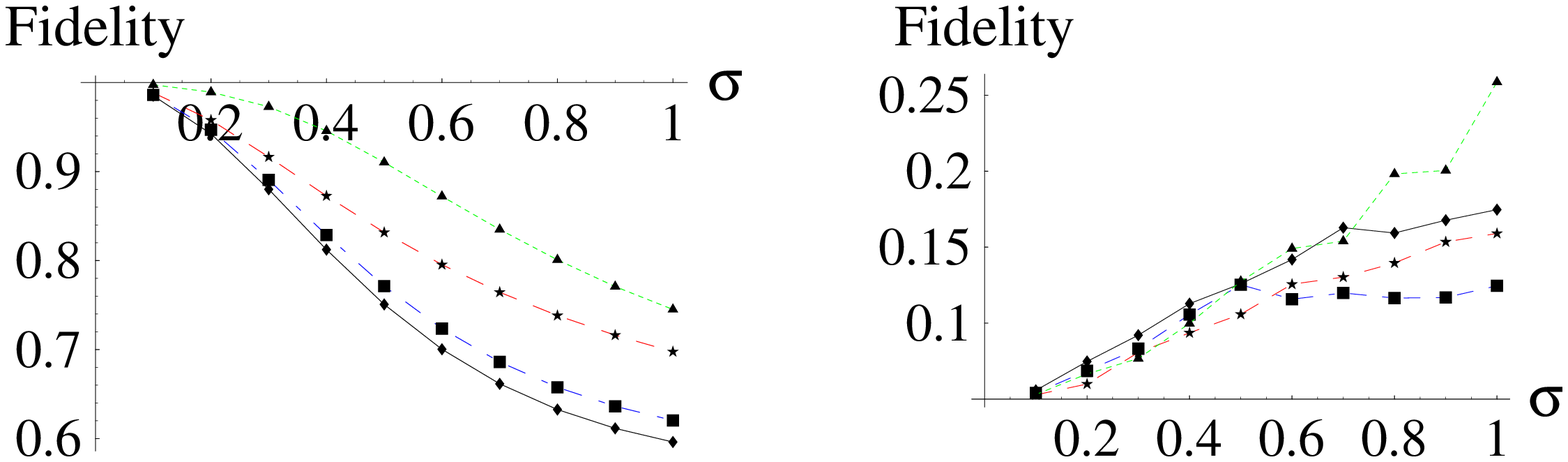,width=9.5cm,height=3.3cm}}
\caption{{\bf (a)}: The fidelity of one-qubit rotations decomposed using the Euler angles $(\zeta,\,\nu,\,\xi)$, when our model of noise is considered. The fidelity is plotted against a common standard deviation $\sigma$ on the unwanted phases for various Euler angles. We have considered $\zeta=\pi/4,\,\nu=\xi=0$ ($\blacklozenge$, solid line), $\zeta=\nu=\pi/2,\,\xi=0$ ($\bigstar$, dashed line), $\zeta=0,\,\nu=\xi=\pi/4$ ($\blacksquare$, dot-dashed line) and $\xi=\zeta=0,\,\nu=\pi$ ($\blacktriangle$, dotted line); {\bf (b)}: The difference $F_{R5}-F_{R7}$ between the gate fidelity of the configurations in Figs.~\ref{genrot} {\bf (a)} and {\bf (b)}.}
\label{fidelityrot}
\end{figure}

The approach used to simulate arbitrary rotations $U_{R}\in SU(2)$~\cite{RBH} requires the decomposition of the rotation in terms of the elements of the Euler-angle vector ${\bm\Omega}=(\zeta,\nu,\xi)$ as $U_{R}=R_x^{\zeta}R_z^{\nu}R_x^{\xi}$, where the elementary rotations around the $j$-axis ($j=x,z$) are given by 
$R_j^{\Omega_i}=\exp\left(-i\Omega_i{\sigma_j}/{2}\right)$.
In order to simulate the action of $U_R$ on the input state $|\psi_{\rm in} \rangle$, we require a five-qubit linear cluster state and the pattern shown in Fig.~\ref{genrot} {\bf (a)}~\cite{RBH}. The pattern consists of single-qubit measurements along directions in direct relation to the Euler angles. These relations can be seen very easily if one pictures Fig.~\ref{genrot} {\bf (a)} as a concatenation of four $BBB_1$'s. In this case we have
\begin{eqnarray}
&&\sigma_x^{s^{xy(\gamma)}_4}{\sf H}R_z^{-\gamma}
\sigma_x^{s^{xy(\beta)}_3}{\sf H}R_z^{-\beta}
\sigma_x^{s^{xy(\alpha)}_2}{\sf H}R_z^{-\alpha}
\sigma_x^{s^{xy(0)}_1}{\sf H} \nonumber \\
&&\equiv\sigma_x^{s^{xy(\gamma)}_4}\sigma_z^{s^{xy(\beta)}_3}\sigma_x^{s^{xy(\alpha)}_2}\sigma_z^{s^{xy(0)}_1}{\sf H}R_z^{-\gamma'}{\sf H}R_z^{-\beta'}{\sf H}R_z^{-\alpha'}{\sf H}\nonumber \\
&&\equiv \tilde{U}_{\Sigma}R_x^{-\gamma'}R_z^{-\beta'}R_x^{-\alpha'},
\end{eqnarray}
where
$\alpha'=(-1)^{s^{xy(0)}_1}\alpha,\,\beta'=(-1)^{s^{xy(\alpha)}_2}\beta$ and $\gamma'=(-1)^{s^{xy(0)}_1 \oplus s^{xy(\beta)}_3}\gamma$ with $\tilde{U}_{\Sigma}^{\dag}= \sigma_z^{s^{xy(0)}_1 \oplus s^{xy(\beta)}_3} \sigma_x^{s^{xy(\alpha)}_2 \oplus s^{xy(\gamma)}_4}$. The measurement bases have now become adaptive, however this does not pose a problem as long as we measure in the order of the qubits in the cluster. Then by redefining the angles for the measurement bases as $\alpha=(-1)^{s^{xy(0)}_1}(-\xi),\,\beta=(-1)^{s^{xy(\alpha)}_2}(-\nu)$ and $\gamma=(-1)^{s^{xy(0)}_1 \oplus s^{xy(\beta)}_3}(-\zeta)$, we simulate the correct rotation $U_{R}$.
The procedure is now used to simulate the rotation of the input state $|\psi_{\rm in} \rangle = a |0 \rangle + b |1 \rangle$,~ ($b=\sqrt{1-a^2}$) via a five-qubit noisy linear cluster. We assume the postselection of the measurement results such that we may retain only the case corresponding to the measurement of the tensorial product $\ket{+}_{1}\ket{+}^{\alpha}_{2}\ket{+}^{\beta}_{3}\ket{+}^{\gamma}_{4}$.
It is easy to write the form of the input state-encoded cluster state 
\begin{equation}
\label{start3}
a\ket{0}_{1}(\ket{\eta}^D+\ket{\mu}^D)_{2,3,4,5}+b\ket{1}_{1}(\ket{\eta}^D-e^{i\theta_{1}}\ket{\mu}^D)_{2,3,4,5},
\end{equation}
where $\ket{\eta}^D$ ($\ket{\mu}^D$) is the part of the noisy subcluster state which has the first qubit in $\ket{0}$ ($\ket{1}$) and the qubit labels have been explicitly introduced. The measurement pattern in Fig.~\ref{genrot} {\bf (a)} together with the specified assumption about the set of outcomes, leads directly to the final state present on qubit $5$ 
\begin{equation}
\label{final3}
\begin{split}
&\ket{\psi_{\rm out}}=\left[a\left(1+e^{i\xi}+e^{i\nu}-e^{i(\xi+\nu+\theta_{2})}\right)+\sqrt{1-a^2}\left(1+e^{i\nu}\right.\right.\\&\left.\left.-e^{i(\xi+\theta_{1})}+e^{i(\xi+\nu+\theta_{1}+\theta_{2})}\right)\right]\ket{+}_{5}+\left[a\left(1+e^{i\xi}-e^{i(\nu+\theta_{3})}\right.\right.\\&\left.+e^{i(\xi+\nu+\theta_{2}+\theta_{3})}\right)+\sqrt{1-a^2}\left(1-e^{i(\xi+\theta_{1})}-e^{i\nu+\theta_{3}}\right.\\&\left.\left.-e^{i(\xi+\nu+\theta_{1}+\theta_{2}+\theta_{3})}\right)\right]e^{i\zeta}(\ket{0}-e^{i\theta_{4}}\ket{1})_{5}.
\end{split}
\end{equation}
This should be compared to the ideally rotated state $U_R\ket{\psi_{\rm in}}$.
The fidelity $F_{R5}=|\bra{\psi_{\rm out}}U_R\ket{\psi_{\rm in}}|^2$, averaged over any possible input state and over individual Gaussian distributions associated with each unwanted $\theta_j$, is shown in Fig.~\ref{fidelityrot} {\bf (a)} for different values of the Euler angles (we label this average benchmark as $\bar{F}_{R5}$). The initial state dependences are washed out by assuming a uniform distribution for each value of $a$, while the Gaussians are all centered on $\theta_j=0$ and have equal standard deviation $\sigma$. It is evident that the average gate fidelity has a considerable dependence on the particular rotation we want to simulate. In this case however, an average over all the rotation angles is meaningless, as the choice of the elements of the vector ${\bm \Omega}$ is imposed by the specific computation protocol desired. The general trend, irrespective of the Euler angles, is that the gate fidelity is reduced by the increase of randomness in the system given by the value of $\sigma$. The reduction can be quite considerable even within {\it moderate} deviations from the ideal values $\theta_j=0,\,\forall{j}$ (see the $\blacksquare$ or the $\blacklozenge$ case in Fig.~\ref{fidelityrot} {\bf (a)} for example, which suffer an average fidelity reduction of about $20\%$, for $\sigma\simeq{0.4}$).

However, this kind of analysis is unable to highlight the effect of redundant qubits initially present in the physical configuration of a cluster. These must be {\it removed} before the effective gate simulation is performed~\cite{RBH}. Such a removal of qubits, not necessary for the simulation of a particular gate, can be achieved by measuring them in the $\sigma_x$ or $\sigma_z$ eigenbasis. For the case of measurements in the $\sigma_z$ eigenbasis, we break any entanglement between that qubit and the rest of the cluster. Whereas for measurements in the $\sigma_x$ eigenbasis, when pairs of qubits are measured, we obtain a reduced cluster state where the qubits no longer affect the QIP protocol being simulated~\cite{RBH}. This removal is a very important point and the influence of noise on QIP protocols has only been partially clarified by our study about quantum state transfer. 
Indeed, a measurement of a qubit in a state affected by a set of $\theta_j$'s, spreads noise throughout the cluster. More precisely, the unwanted phase $\theta_j$ which was attached to the $j-$th qubit, eliminated from the cluster by a measurement, is then {\it inherited} by the surviving qubits surrounding the $j$-th one. This leaves us with a smaller cluster state which is plagued by a larger number of unwanted phases. This {\it noise inheritance effect} must obviously be kept to a minimum. 

Moreover, this effect does not depend on the particular angle chosen for the measurement basis and noise inheritance appears after any measurement (belonging to a legitimate measurement pattern) on a cluster state. It thus becomes an intrinsic feature of the inherently noisy cluster state generation we are addressing~\cite{noiarchivio}. A way to give an explicit account of the inheritance effect is by fictitiously modifying the configuration in Fig.~\ref{genrot} {\bf (a)} as shown in panel {\bf (b)}, where qubits $3$ and $6$ must be considered as redundant. After their elimination, via measurements in the $\sigma_x$-eigenbasis (which do not break the channels between the surviving qubits $2~\&~4$ and $5~\&~7$), the physical layout is exactly the five-qubit linear cluster considered in panel {\bf (a)}. We have calculated the gate fidelity $F_{R7}$ after the removal of these qubits and the correction of the resulting cluster state via local operations, as if a Hadamard gate has been performed between qubits $3~\&~4$ and $6~\&~7$. Analyzing the effect of noise spreading through the measurements, we find that the average of $F_{R7}$ over the input states and individual Gaussian distributions ($\bar{F}_{R7}$) is always smaller than $\bar{F}_{R5}$. This is shown in Fig.~\ref{fidelityrot} {\bf (b)}, where we have plotted the difference $\bar{F}_{R5}-\bar{F}_{R7}$ (always positive) against $\sigma$, for the rotation angles considered in Fig.~\ref{fidelityrot} {\bf (a)}. The differences between the two cluster configurations can easily exceed $10\%$ and are larger for increased randomness in the cluster. Their behaviors as the Euler angles are changed, are almost uniform until $\sigma\simeq0.6$. Then for increased randomness, the specific way in which the inherited phases are distributed within the structure of the cluster state becomes relevant and some discrepancies occur. The message here is that there is a counterintuitive dependence of the gate fidelity on the specific angles of rotation: some rotations are more {\it exposed} to our model of noise than others.                                                                                
\subsection{{\sf CNOT} gate}
\label{cnot}

In order to complete our analysis of QIP protocols in the \QC~scheme with intrinsic noise, we address the simulation of a {\sf CNOT} as the paradigmatic example of an entangling two-qubit gate~\cite{nielsenchuang}. We first examine the behavior of the original proposal~\cite{RBH} under noisy conditions and in the presence of a redundant qubit. Then, we will compare it with other configurations for {\sf CNOT} simulation, where the resource requirements can be dramatically reduced down to no more than four qubits~\cite{noiarchivio}. 

\begin{figure}[ht]
\centerline{{\bf (a)}}
\begin{minipage}[c]{0.1\textwidth}
\centerline{\psfig{figure=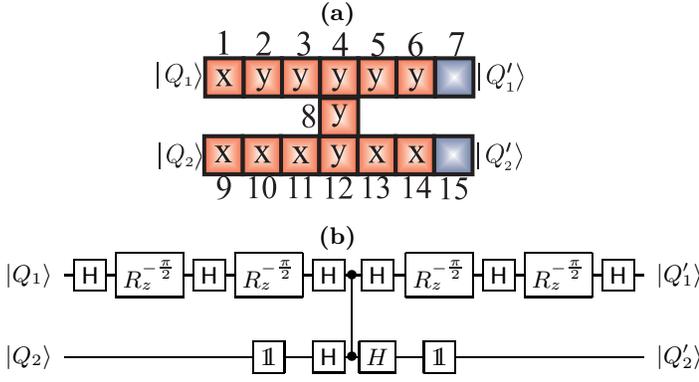,width=5.0cm,height=2.3cm}}
\end{minipage}\\
\vspace*{0.3cm}
\centerline{{\bf (b)}}
\hskip-3.0cm
\begin{minipage}[l]{0.25\textwidth}
\mbox{
\Qcircuit @C=0.4em @R=1.8em {
  \lstick{\ket{Q_1}} & \gate{\sf H}&\gate{R^{-\frac{\pi}{2}}_{z}} & \gate{\sf H}&\gate{R^{-\frac{\pi}{2}}_{z}}&\gate{\sf H}& \ctrl{1} & \gate{\sf H}&\gate{R^{-\frac{\pi}{2}}_{z}} & \gate{\sf H}&\gate{R^{-\frac{\pi}{2}}_{z}}&\gate{\sf H}& \rstick{\ket{Q_1'}} \qw \\
  \lstick{\ket{Q_2}} & \qw &\qw&\qw&\gate{\openone}& \gate{\sf H}& \ctrl{-1} & \gate{H}& \gate{\openone}&\qw &\qw&\qw&\rstick{\ket{Q_2'}} \qw }}
\end{minipage}
\centerline{}
\caption{{\bf (a)}: {\it Squashed-}I configuration for a $\sf CNOT$ simulation. The input (output) control and target logical qubit are qubits $1$ ($7$) and $9$ ($15$) respectively, with qubit $8$ as a bridging qubit. {\bf (b)}: The equivalent quantum circuit as a concatenation of $BBB$'s and $EBB$'s.}
\label{cnotBBB}
\end{figure}

The input states will be denoted $|Q_{1} \rangle=a |0 \rangle + b |1 \rangle$ and $|Q_{2} \rangle=c |0 \rangle + d |1 \rangle$ which encode the control and target state respectively. We consider the fifteen-qubit bidimensional cluster state (the {\it squashed}-I configuration) whose layout and measurement pattern are shown in Fig.~(\ref{cnotBBB}) {\bf (a)}. Its workings in terms of stabilizer formalism can be found in~\cite{RBH}. Here, we are interested in the concatenation technique, which gives an immediate picture of the equivalent quantum gates simulated by this cluster configuration. Using the $BBB$'s and $EBB$'s introduced in Section~\ref{tessellation}, it is straightforward to derive the equivalent quantum circuit in Fig.~\ref{cnotBBB} {\bf (b)}. The role played by the $EBB$ introduced in Fig.~\ref{fig:ygate}, bridging the two otherwise independent subclusters, is crucial here. 
The byproduct operator can be evaluated starting from this equivalent decomposition and explicitly found to be
$U_{\Sigma}=\otimes_{j=7,15}\sigma_z^{\gamma_z^{(j)}}\sigma_x^{\gamma_x^{(j)}}$
with
\begin{equation}
\begin{split}
\gamma_x^{(7)}&={s}_2^{y}+{s}_3^{y}+{s}_5^{y}+{s}_6^{y},\\
\gamma_x^{(15)}&={s}_2^{y}+{s}_3^{y}+{s}_8^{y}+{s}_{10}^{x}+{s}_{12}^{y}+{s}_{14}^{x},\\
\gamma_z^{(7)}&={s}_1^{x}+{s}_3^{y}+{s}_4^{y}+{s}_5^{y}+{s}_8^{y}+{s}_9^{x}+{s}_{11}^{x}+1,\\
\gamma_z^{(15)}&={s}_9^{x}+{s}_{11}^{x}+{s}_{13}^{x},
\end{split}
\end{equation}
where $s^{y}_i=s^{xy(\pi/2)}_i$ and $s^{x}_i=s^{xy(0)}_i$ irrespective of $i$.
We now consider the simulation of a {\sf CNOT} with a noisy squashed-I cluster. 
The technique already exploited in Section~\ref{rotazioni} in order to write an implicit (but manageable) expression for the global state of a cluster in a given configuration is used again and the starting point for our calculations is the state
\begin{equation}
\label{start}
\ket{\varphi}^D_{1\rightarrow4}\otimes\ket{\varphi}^D_{5\rightarrow7}\otimes\ket{+}_{8}\otimes\ket{\varphi}^D_{9\rightarrow12}\otimes\ket{\varphi}^D_{13\rightarrow15}.
\end{equation}
Here $\ket{\varphi}^D=\ket{\psi}^D+\ket{\chi}^D=\ket{\eta}^D+\ket{\mu}^D$, where $\ket{\psi}^D$ is the part of $\ket{\varphi}^D$ which has the last qubit in ${\ket{0}}$, while $\ket{\chi}^D$ is the part with the last qubit in $\ket{1}$. The explicit form of $\ket{\psi}^D,\,\ket{\chi}^D,\,\ket{\eta}^{D}$ and $\ket{\mu}^D$ is given in Tab.~\ref{notazionecnot}. In Eq.~(\ref{start}) the encoded logical input state of the control and target qubits, as well as the state of the bridging qubit $8$, have been properly singled out. The $1\rightarrow4$ ($9\rightarrow12$) symbol means that qubits $1,2,3,4$ ($9,10,11,12$) are involved. By applying the entangling operations $S^{4,8}_D$, $S^{8,12}_D$, $S^{4,5}_D$ and $S^{12,13}_D$, the state of the cluster becomes
\begin{equation}
\label{final}
\begin{split}
&\left[\ket{\psi}^D_{1\rightarrow4}\ket{\varphi}^D_{5\rightarrow7}+\ket{\chi}^D_{1\rightarrow4}\left(\ket{\eta}^D-e^{i\theta^R_{4}}\ket{\mu}^D\right)_{5\rightarrow7}\right]\ket{0}_{8}\otimes\\&
\left[\ket{\psi}^D_{9\rightarrow12}\ket{\varphi}^D_{13\rightarrow15}+\ket{\chi}^D_{9\rightarrow12}\left(\ket{\eta}^D-e^{i\theta_{12}}\ket{\mu}^D\right)_{13\rightarrow15}\right]+\\
&\left[\ket{\psi}^D_{1\rightarrow4}\ket{\varphi}^D_{5\rightarrow7}-e^{i\theta^{C}_{4}}\ket{\chi}^D_{1\rightarrow4}\left(\ket{\eta}^D-e^{i\theta^R_{4}}\ket{\mu}^D\right)_{5\rightarrow7}\right]\ket{1}_{8}\otimes\\
&\left[\ket{\psi}^D_{9\rightarrow12}\ket{\varphi}^D_{13\rightarrow15}-e^{i\theta^{}_{8}}\ket{\chi}^D_{9\rightarrow12}\left(\ket{\eta}^D-e^{i\theta_{12}}\ket{\mu}^D\right)_{13\rightarrow15}\right].
\end{split}   
\end{equation}
Here, $\theta^{R}_{4}$ ($\theta^{C}_{4}$) denotes the unwanted phase introduced when the entanglement between qubits $4$ and $5$ (qubits $4$ and $8$) is considered. 
Using this notation, the entangling operation which glues together two subclusters is considerably simplified. For instance, gluing $\ket{\varphi}^D_{1\rightarrow4}$ and $\ket{\varphi}^D_{5\rightarrow{7}}$ leads to
\begin{eqnarray}
\label{esempio}
&&S^{4,5}_{D}\ket{\varphi}^D_{1\rightarrow4}\otimes\ket{\varphi}^D_{5\rightarrow7}=S^{4,5}_{D}(\ket{\psi}^D+\ket{\chi}^D)_{1\rightarrow4} \otimes \nonumber \\
&&(\ket{\eta}^D+\ket{\mu}^D)_{5\rightarrow7}=\ket{\psi}^D_{1\rightarrow4}\ket{\varphi}^D_{5\rightarrow7}+\ket{\chi}^D_{1\rightarrow4}\ket{\eta}^D_{5\rightarrow7}\nonumber \\
&&-e^{i\theta_{4}}\ket{\chi}^D_{1\rightarrow4}\ket{\mu}^D_{5\rightarrow7},
\end{eqnarray}
which is a seven-qubit linear cluster state. It should be noted that this expression has been obtained without actually writing the subcluster states in the computational basis. This method is therefore space-saving and computationally useful. On the other hand, by dealing each time with small subcluster states, it is handy to find out the explicit form of the state of the output logical qubits after the application of the appropriate measurement pattern. 
It is easy to build up a table for the transformations occurring after the completed measurements. 
\begin{figure}[b]
\hspace*{0.5cm}\psfig{figure=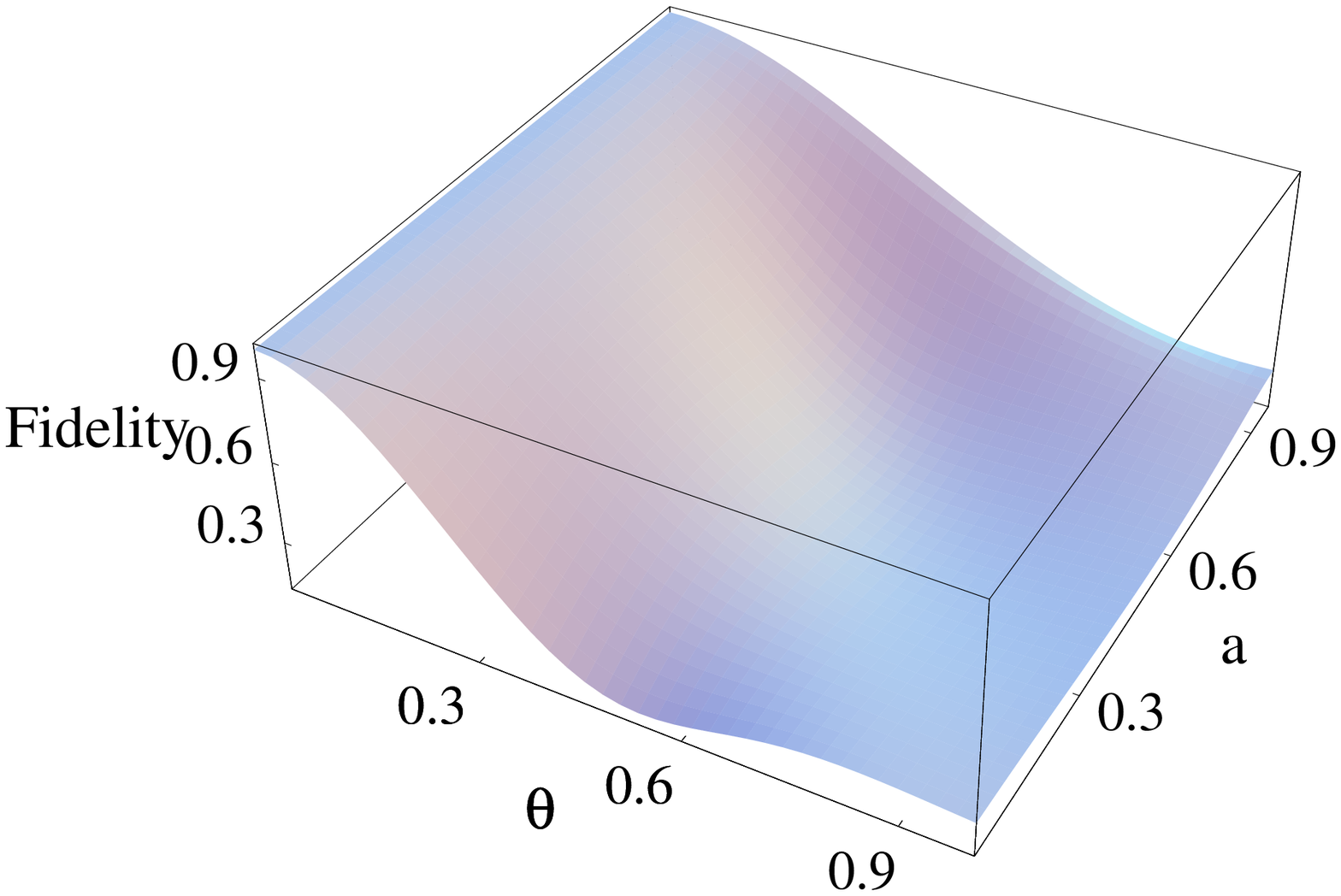,width=4.0cm,height=3.3cm}
\psfig{figure=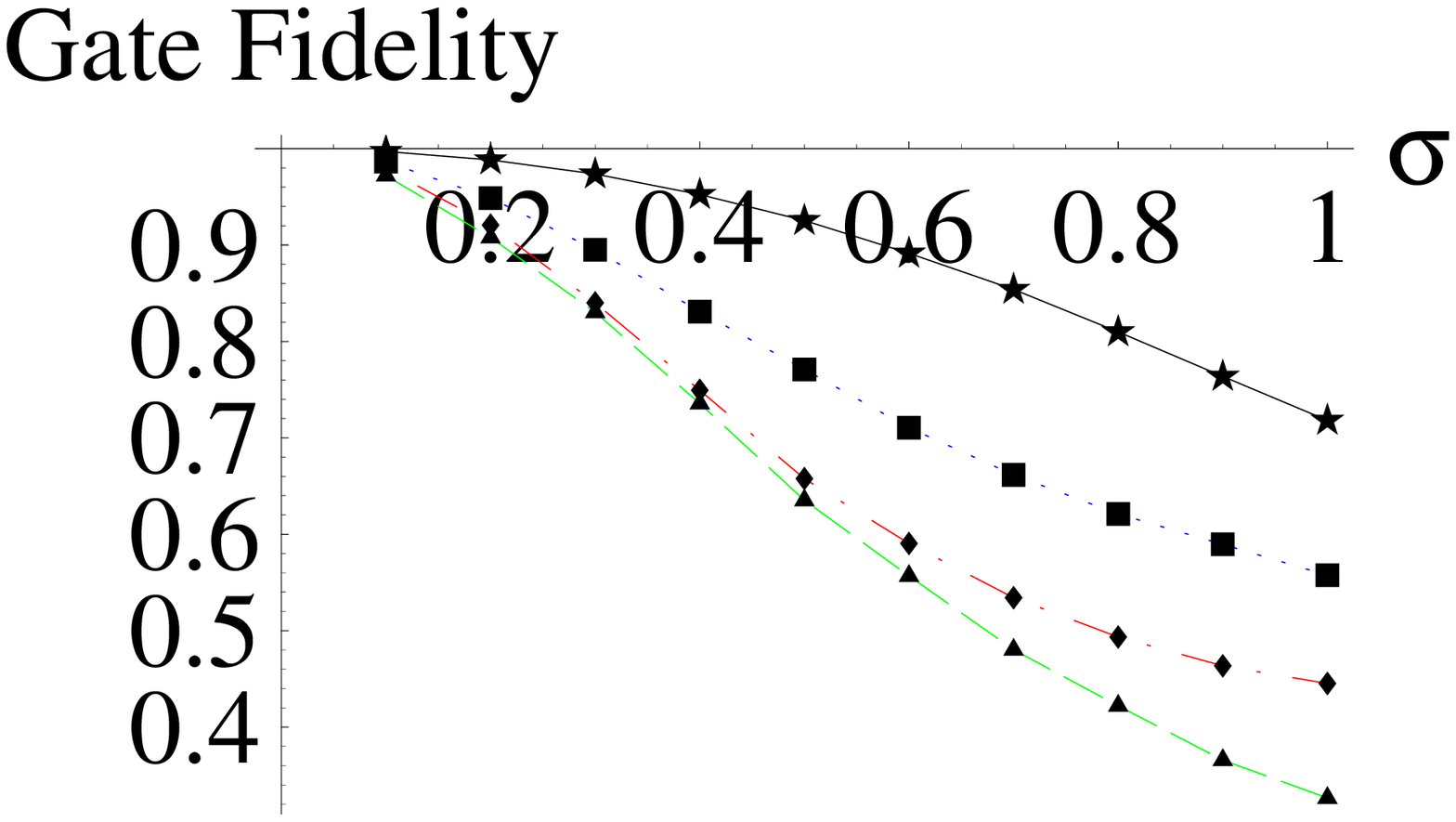,width=4.0cm,height=3.0cm}
\caption{{\bf (a)}: Fidelity for a squashed-I $\sf CNOT$ plotted against the unwanted phase $\theta$ and the input control-state coefficient $a$. In this plot $a=c$. {\bf (b)}: Gate fidelities for different simulations of a {\sf CNOT}. From top to bottom curve, we show the fidelity of the four-qubit {\sf CNOT} of Fig.~\ref{fig:tfig8}, the helix configuration, the squashed-I and the squashed-I with an additional bridging qubit.}
\label{fig:tfigcnotdog}
\end{figure}
\begin{table} [t]
\begin{ruledtabular}
\begin{tabular}{|l|c|}\hline
\hskip0.cm State& Explicit form in the computational basis\\ \hline\hline
$\hskip0.cm\ket{\psi}^D_{1\rightarrow4}$&$(a\ket{0}+b\ket{1})_1(\ket{000}+\ket{010})_{2\rightarrow4}$\\&
$+(a\ket{0}-e^{i\theta_{1}}b\ket{1})_1(\ket{100}-e^{i\theta_{2}}\ket{110})_{2\rightarrow4}$\\ \hline
$\hskip0.cm\ket{\chi}^D_{1\rightarrow4}$&$(a\ket{0}+b\ket{1})_1(\ket{001}-e^{i\theta_{3}}\ket{011})_{2\rightarrow4}+$\\&
$(a\ket{0}-e^{i\theta_{1}}b\ket{1})_1(\ket{101}+e^{i(\theta_{2}+\theta_{3})}\ket{111})_{2\rightarrow4}$\\ \hline
$\hskip0.cm\ket{\eta}^D_{5\rightarrow7}$& $(\ket{000}+\ket{010}+\ket{001}-e^{i\theta_{6}}\ket{011})_{5\rightarrow7}$  \\ \hline
$\hskip0.cm\ket{\mu}^D_{5\rightarrow7}$&$(\ket{100}+\ket{101}-e^{i\theta_{5}}\ket{110}+e^{i(\theta_{6}+\theta_{5})}\ket{111})_{5\rightarrow7}$  \\ \hline
$\hskip0.cm\ket{\psi}^D_{9\rightarrow12}$&$(c\ket{0}+d\ket{1})_9(\ket{000}+\ket{010})_{10\rightarrow12}+$\\&
$(c\ket{0}-e^{i\theta_{9}}d\ket{1})_9(\ket{100}-e^{i\theta_{10}}\ket{110})_{10\rightarrow12}$\\ \hline
$\hskip0.cm\ket{\chi}^D_{9\rightarrow12}$&$(c\ket{0}+d\ket{1})_9(\ket{001}-e^{i\theta_{11}}\ket{011})_{10\rightarrow12}+$\\&
$(c\ket{0}-e^{i\theta_{9}}d\ket{1})_9(\ket{101}+e^{i(\theta_{10}+\theta_{11})}\ket{111})_{10\rightarrow12}$\\ \hline
$\hskip0.cm\ket{\eta}^D_{13\rightarrow15}$& $(\ket{000}+\ket{010}+\ket{001}-e^{i\theta_{14}}\ket{011})_{13\rightarrow15}$  \\ \hline
$\hskip0.cm\ket{\mu}^D_{13\rightarrow15}$&$(\ket{100}+\ket{101}-e^{i\theta_{13}}\ket{110}+e^{i(\theta_{13}+\theta_{14})}\ket{111})_{13\rightarrow15}$  \\ \hline
\end{tabular}
\end{ruledtabular}
\caption{The notation used in Eq.~(\ref{final}) for the noisy squashed-I cluster state used to simulate a $\sf CNOT$.\label{notazionecnot}}
\end{table}
Following these lines, the gate fidelity $F_{\sf CNOT}$ can be explicitly evaluated by hand. A plot is given in Fig.~\ref{fig:tfigcnotdog} {\bf (a)} against both $\theta_{j}=\theta$ ($\forall j$ involved in Eq.~(\ref{final})) and $a=c$ (for convenience) with the normalization conditions $b=\sqrt{1-a^2},\,d=\sqrt{1-c^2}$.
As in the case of single-qubit rotations, we have considered the post-selection of the event corresponding to $s^{y}_{i}=s^{x}_{j}=0$ among the set of outcomes resulting from the measurements. 
In this case, the decoding operator is $\tilde{U}_{\Sigma}^{\dag}(\{0\})=\sigma_{z}^{(7)}\otimes\openone^{(15)}$. Behaviors qualitatively similar to Fig.~\ref{fig:tfigcnotdog} {\bf (a)} can be observed for any other choice of the relation between $a$ and $c$ with this plot having the merit of showing an almost uniform behavior of $F_{\sf CNOT}$ against $a$, for a fixed $\theta$ and a fast decay of the gate fidelity is found for non-zero values of the unwanted phases. Near $\theta=0.6$ and $a=c=0.5$, $F_{\sf CNOT}\simeq{0.2}$ is found and is never exceeded for fixed $a$, whatever the choice for the relation between $a$ and $c$. By showing the fidelity behavior in Fig.~\ref{fig:tfigcnotdog} {\bf (a)}, we produce a significant example of the performances of the squashed-I {\sf CNOT} simulation in the presence of our model of noise, showing that the one-way model has to face a considerable decay in the two-qubit gate fidelity. These conclusions are strengthened by the calculation of the average fidelity over Gaussian distributions for the unwanted phases shown in the dot-dashed line ($\blacklozenge$ symbol) in Fig.~\ref{fig:tfigcnotdog} {\bf (b)}. Although a more complete analysis requires an average over all the outcome configurations~\cite{limite}, this example provides sufficient physical insight.

We can study the way in which the noise inheritance attacks the $\sf CNOT$ fidelity by analyzing the simple but interesting example of a modified squashed-I 
where the bridging zone between the control and target subclusters includes an additional redundant qubit.
We remove this qubit from the cluster (via a measurement in the $\sigma_x$-eigenbasis) retrieving the fifteen-qubit cluster state through appropriate local operations on qubit $12$. 
Following the lines depicted above {\it i.e.} attaching Gaussian distributions of standard deviation $\sigma$ to the unwanted phases appearing in the noisy cluster state, 
we can verify that the average gate fidelity is spoiled, as shown in Fig.~\ref{fig:tfigcnotdog} {\bf (b)} ($\blacktriangle$, dashed line) from the expected spread of additional phases relative to the removed qubit~\cite{noiarchivio}. 

\subsection{Alternative routes to $\sf CNOT$}

Our discussion about noise inheritance reinforces the view we introduced previously, concerning the importance of keeping the number of qubits in a cluster as low as possible. The analysis of the squashed-I layout revealed it to be rather prone to the effect of the intrinsic noise model. We ascribe this to the {\it expensive} nature of the configuration in terms of the number of qubits in the cluster. Thus there is a necessity for looking at different ways in which an entangling two-qubit gate can be simulated through cluster states. Here, we would like to provide an example of alternative cluster configurations which are able to simulate a $\sf CNOT$ gate involving less qubits than the squashed-I. 

\begin{figure}[t]
\hskip-1.0cm{\bf (a)}\hspace*{3.8cm}{\bf (b)}\\
\hskip0.7cm
\begin{minipage}[c]{0.1\textwidth}
\centerline{\psfig{figure=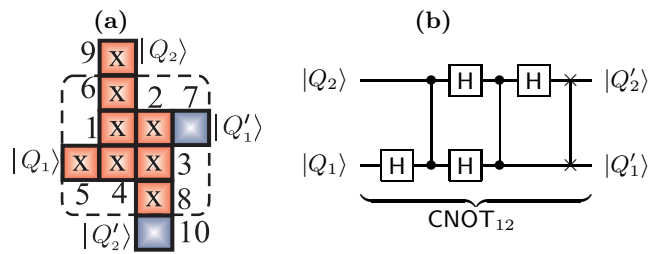,width=3.4cm,height=2.9cm}}
\end{minipage}%
\hskip1.8cm
\begin{minipage}[l]{0.2\textwidth}
\vspace*{0cm}
\mbox{
 \Qcircuit @C=0.85em @R=2.2em {
  \lstick{\ket{Q_2}}&\qw&\ctrl{1}&\gate{\sf H}&\ctrl{1}&\gate{\sf H}&\qswap&\rstick{\ket{Q_2'}}\qw\\
\lstick{\ket{Q_1}}&\gate{\sf H}&\ctrl{-1}&\gate{\sf H}&\ctrl{-1}&\qw&\qswap\qwx&\rstick{\ket{Q_1'}}\qw\gategroup{1}{2}{2}{7}{1.6em}{_\}}}}
\\
\vspace*{0.1cm}
{${\sf CNOT}_{12}$}
\end{minipage}
\caption{{\bf (a)}: {\it Helix} layout and measurement pattern for a $\sf CNOT$ simulated through a $\sf C_{\pi}PHASE$ gate (within the dashed square) and two Hadamard gates involving qubits $6~\&~9$ and $8~\&~10$.{\bf (b)}: Equivalent quantum circuit drawn by exploiting the concatenation of eight $BBB_1$'s and one box-shaped $EBB$ involving qubits $1,2,3,4$ (see Fig. {\ref{fig:tfig7}}).}
\label{cnotcphase}
\end{figure}

We consider the qubit layout and measurement pattern sketched in Fig.~\ref{cnotcphase} {\bf (a)}, which simulates a $\sf CNOT$ using ten qubits. We will refer to this configuration as the {\it helix} configuration. The scheme is based on the simulation of a $\sf C_{\pi}PHASE$ gate~\cite{nielsenchuang} (within the dashed box) and realizes the transformation ${\sf H}({{\sf C_{\pi}PHASE}}){\sf H}\equiv{\sf CNOT}$. A more detailed analysis of the gate simulation, using the concatenation technique is given in Fig.~\ref{cnotcphase} {\bf (b)}, where a relabelling of the logical output qubits, effectively equivalent to a {\sf SWAP} gate is required. The crucial feature is the use of a box cluster (see Fig.~\ref{fig:tfig7}). 

It is possible to carry out an analytic calculation of the dynamics of a noisy helix cluster state using the same technique highlighted previously. The resulting average gate fidelity is shown in Fig.~\ref{fig:tfigcnotdog} {\bf (b)} ($\blacksquare$, dotted line).
As before, we assume the post-selection of the event corresponding to the set of outcomes $s^{x}_{i}=0$ ({\it i.e.} all the measured qubits are found in $\ket{+}$). This results in no decoding operator  at the end of the procedure, a clear advantage with respect to the squashed-I $\sf CNOT$. The usual individual Gaussians have been considered, with $\sigma$ as their standard deviation
and a noticeable improvement in the gate fidelity is observed, compared to the squashed-I configuration.

The situation can be further improved by looking at the squashed-I cluster configuration and examining more closely the simulation performed there. 
It is straightforward to recognize that the non-local nature of the {\sf CNOT} gate in Fig.~\ref{cnotBBB} {\bf (b)} is all in the ${\sf C_{\pi}PHASE}$ gate sandwiched by the Hadamards on the $\ket{Q_2}-\ket{Q'_2}$ (target) line. The remainder of the circuit realizes local operations on the control qubit, which are unnecessary for the {\sf CNOT} simulation. Thus, there is a consistent {\it redundancy} in this cluster configuration. Stripping the squashed-I cluster bare to eliminate the unnecessary local parts, leads directly to the $EBB$ already introduced in Fig.~\ref{fig:ygate} at its very core. But even this is unnecessary as we already know that it is possible to obtain a better configuration which provides an even more economical configuration for a {\sf CNOT} simulation. This is the four-qubit $EBB$ discussed in Section~\ref{tessellation} Fig.~\ref{fig:tfig8}, which naturally simulates a {\sf CNOT} (with output states in the $\sigma_x$-eigenbasis). Our aim is therefore confronting the performance of the noisy version of this simple configuration with the other simulations treated so far. The results are shown in Fig.~\ref{fig:tfigcnotdog} {\bf (b)} ($\bigstar$ and solid line).  Evidently the gate fidelity corresponding to this linear layout is vastly superior to any other case treated so far~\cite{noiarchivio}. This four-qubit $EBB$ {\sf CNOT} is important for two essential reasons. First, it can be seen as the ultimate confirmation that in a noisy scenario, those features which the ideal one-way model takes for granted, {\it i.e.} the management of an arbitrarily large cluster and the innocuousness of the measurements performed in order to process the encoded information, sensibly affect a computational task when imperfections are embedded in the cluster state. Second, it is evident that more economical configurations for gate simulation are required if reliable computation is to be performed. The four-qubit linear $EBB$ actually provides an economical {\sf CNOT} simulation in which the number of parameters involved corresponds to the number of qubits in the cluster resource. These points, already extensively commented in~\cite{noiarchivio}, have been enriched here by a more detailed analysis. 
\section{Remarks}
\label{remarks}

We have addressed several very important points in the cluster state model for QC. The construction and analysis of cluster configurations suitable for the simulation of desired quantum circuits, has been simplified by the introduction of a class of elementary $BBB$'s. We have seen how they can be concatenated together in order to model even complicated circuit configurations. Our strategy allows one to immediately figure out the form of the decoding operator in order to eliminate the effects of the measurement-induced randomness in a simulation. We believe this represents a valuable tool for the purposes of configuration-design, which is helpful both theoretically and experimentally. Indeed, by properly designing the cluster state resource, it is possible to minimize the number of redundant qubits in a circuit simulation. This point is related to the second task accomplished by this paper, namely the detailed analysis of QIP using intrinsic imperfections in the generation of cluster configurations. We have shown that uncontrollable randomness in the qubit-qubit interactions which create a cluster state, affect both communication and computation protocols based on the one-way model. A direct consequence of our analysis is that in the processing of information encoded in a cluster state, both the number of qubits involved and the measurements to be performed must be carefully managed. Our study paves the way toward research of gate simulations performed by using only small clusters of just a few qubits. The existence of such economic configurations has been further commented here. 
\acknowledgments

We acknowledge support by UK EPSRC, KRF (2003-070-C00024),  DEL and IRCEP. 



\begin{thebibliography}{99}
\bibitem{clifford} For example {{\sf CNOT}, {\sf H} (Hadamard) and $\pi/2$-phase gates are in the Clifford group.}
\bibitem{commentoent} The class of graph-states is larger than the class of cluster states and contains the latter~\cite{rb1,rb2}.
\bibitem{limite} A calculation of the complete set of outcomes is prohibitive. Indeed, $N=10$ represents a sort of limiting value on the dimension of a cluster, beyond which the computational time becomes exeedingly large.
\end{thebibliography}
\end{document}